\documentclass[usenatbib]{mnras}
\usepackage[T1]{fontenc}
\DeclareRobustCommand{\VAN}[3]{#2}
\let\VANthebibliography\thebibliography
\def\thebibliography{\DeclareRobustCommand{\VAN}[3]{##3}\VANthebibliography}
\usepackage{graphicx}
\usepackage{amsmath}
\usepackage{amssymb}
\usepackage{color}
\usepackage{url}
\usepackage{dsfont}
\usepackage{subfig}
\usepackage{pdflscape}
\usepackage{changepage}
\usepackage{CJKutf8}

\usepackage{threeparttable}

\newcommand\lsim{\mathrel{\rlap{\lower4pt\hbox{\hskip1pt$\sim$}}
        \raise1pt\hbox{$<$}}}
\newcommand\gsim{\mathrel{\rlap{\lower4pt\hbox{\hskip1pt$\sim$}}
        \raise1pt\hbox{$>$}}}
\newcommand{\lya}{\ifmmode\mathrm{Ly}\alpha\else{}Ly$\alpha$\fi}
\newcommand{\lyb}{\ifmmode\mathrm{Ly}\beta\else{}Ly$\beta$\fi}
\newcommand{\igm}{\ifmmode\mathrm{IGM}\else{}IGM\fi}
\newcommand{\lae}{\ifmmode\mathrm{LAE}\else{}LAE\fi}
\newcommand{\h}{\ifmmode\mathrm{H}\else{}H\fi}
\newcommand{\hi}{\ifmmode\mathrm{H\,{\scriptscriptstyle I}}\else{}H\,{\scriptsize I}\fi}
\newcommand{\hii}{\ifmmode\mathrm{H\,{\scriptscriptstyle II}}\else{}H\,{\scriptsize II}\fi}
\newcommand{\cmb}{\ifmmode\mathrm{CMB}\else{}CMB\fi}
\newcommand{\qso}{\ifmmode\mathrm{QSO}\else{}QSO\fi}
\newcommand{\eor}{\ifmmode\mathrm{EoR}\else{}EoR\fi}
\newcommand{\heii}{\ifmmode\mathrm{He\,{\scriptscriptstyle II}}\else{}He\,{\scriptsize II}\fi}
\newcommand{\heiii}{\ifmmode\mathrm{He\,{\scriptscriptstyle III}}\else{}He\,{\scriptsize III}\fi}
\newcommand{\ciii}{\ifmmode\mathrm{C\,{\scriptscriptstyle III]}}\else{}C\,{\scriptsize III]}\fi}
\newcommand{\oiii}{\ifmmode\mathrm{O\,{\scriptscriptstyle III}}\else{}O\,{\scriptsize III}\fi}
\newcommand{\aliii}{\ifmmode\mathrm{Al\,{\scriptscriptstyle III}}\else{}Al\,{\scriptsize III}\fi}
\newcommand{\mgii}{\ifmmode\mathrm{Mg\,{\scriptscriptstyle II}}\else{}Mg\,{\scriptsize II}\fi}
\newcommand{\fe}{\ifmmode\mathrm{Fe}\else{}Fe\fi}
\newcommand{\nv}{\ifmmode\mathrm{N\,{\scriptscriptstyle V}}\else{}N\,{\scriptsize V}\fi}
\newcommand{\niv}{\ifmmode\mathrm{N\,{\scriptscriptstyle IV]}}\else{}N\,{\scriptsize IV]}\fi}
\newcommand{\cii}{\ifmmode\mathrm{C\,{\scriptscriptstyle II}}\else{}C\,{\scriptsize II}\fi}
\newcommand{\civ}{\ifmmode\mathrm{C\,{\scriptscriptstyle IV}}\else{}C\,{\scriptsize IV}\fi}
\newcommand{\siv}{\ifmmode\mathrm{Si\,{\scriptscriptstyle IV}}\else{}Si\,{\scriptsize IV}\fi}
\newcommand{\siii}{\ifmmode\mathrm{Si\,{\scriptscriptstyle II}}\else{}Si\,{\scriptsize II}\fi}
\newcommand{\siiii}{\ifmmode\mathrm{Si\,{\scriptscriptstyle III]}}\else{}Si\,{\scriptsize III]}\fi}
\newcommand{\ovi}{\ifmmode\mathrm{O\,{\scriptscriptstyle VI}}\else{}O\,{\scriptsize VI}\fi}
\newcommand{\sioiv}{\ifmmode\mathrm{Si\,{\scriptscriptstyle IV}\,\plus O\,{\scriptscriptstyle IV]}}\else{}Si\,{\scriptsize IV}\,+O\,{\scriptsize IV]}\fi}

\usepackage{newtxtext,newtxmath}
\usepackage{tikz}
\usetikzlibrary{shapes.geometric}

\newcommand*\circled[1]{\tikz[baseline=(char.base)]{
            \node[shape=ellipse,draw,inner sep=0.5pt] (char) {#1};}}

\pdfoutput=1

\makeatletter
\providecommand\phantomcaption{\caption@refstepcounter\@captype}
\makeatother

\title[Blind QSO challenge]{Blind QSO reconstruction challenge: Exploring methods to reconstruct the Ly$\alpha{}$ emission line of QSOs}
\author[B. Greig et al.] {Bradley~Greig$^{1,2,3}$\thanks{E-mail:~greigb@unimelb.edu.au}, S. E. I. Bosman$^{4,5}$, F. B. Davies$^5$, 
D. {{\v{D}}urov{\v{c}}{\'\i}kov{\'a}}$^6$, H. Fathivavsari$^7$, \newauthor 
B. Liu$^8$, R. A. Meyer$^{9,5}$, Z. Sun (孙泽昌)$^{10}$, V. D’Odorico$^{11,12,13}$, S. Gallerani$^{12}$, \newauthor 
A. Mesinger$^{12}$ \& Y.-S. Ting (丁源森)$^{2,14,15,16,17}$\\
$^1$School of Physics, University of Melbourne, Parkville, VIC 3010, Australia \\
$^2$Research School of Astronomy \& Astrophysics, Australian National University, Canberra, ACT 2611, Australia \\
$^3$ARC Centre of Excellence for All-Sky Astrophysics in 3 Dimensions (ASTRO 3D) \\
$^4$Institute for Theoretical Physics, Heidelberg University, Philosophenweg 12, D-69120, Heidelberg, Germany \\
$^5$Max-Planck-Institut f{\"u}r Astronomie, K{\"o}nigstuhl 17, D-69117 Heidelberg, Germany \\
$^6$MIT Kavli Institute for Astrophysics and Space Research, 77 Massachusetts Avenue, Cambridge, 02139, Massachusetts, USA \\
$^7$School of Astronomy, Institute for Research in Fundamental Sciences (IPM), P.O. Box 19395-5531, Tehran, Iran \\
$^8$Department of Physics, North Carolina State University, Raleigh, NC 27695, USA \\
$^9$Department of Astronomy, University of Geneva, Chemin Pegasi 51, 1290 Versoix, Switzerland \\
$^{10}$Department of Astronomy, Tsinghua University, Beijing 100084, Peopleʼs Republic of China \\
$^{11}$INAF–Osservatorio Astronomico di Trieste, Via G.B. Tiepolo, 11, I-34143, Trieste, Italy \\
$^{12}$Scuola Normale Superiore, Piazza dei Cavalieri, I-56126 Pisa, Italy \\
$^{13}$IFPU–Institute for Fundamental Physics of the Universe, via Beirut 2, I34151 Trieste, Italy \\
$^{14}$School of Computing, Australian National University, Acton ACT 2601, Australia \\
$^{15}$Department of Astronomy, The Ohio State University, Columbus, OH 45701, USA \\
$^{16}$Center for Cosmology and AstroParticle Physics (CCAPP), The Ohio State University, Columbus, OH 43210, USA \\
$^{17}$Department of Physics, Faculty of Science, Universiti Malaya, 50603 Kuala Lumpur, Malaysia
}
\pubyear{2024}

\begin{document}
\label{firstpage}
\pagerange{\pageref{firstpage}--\pageref{lastpage}}
\begin{CJK}{UTF8}{gkai} 
\maketitle
\end{CJK}
\begin{abstract}
\noindent
Reconstructing the intrinsic \lya{} line flux from high-$z$ QSOs can place constraints on the neutral hydrogen content of the intergalactic medium during reionisation. There are now $\gtrsim10$ different \lya{} reconstruction pipelines using different methodologies to predict the \lya{} line flux from correlations with the spectral information redward of \lya{}. However, there have been few attempts to directly compare the performance of these pipelines. Therefore, we devised a blind QSO challenge to compare these reconstruction pipelines on a uniform set of objects. Each author was provided de-identified, observed rest-frame QSO spectra with spectral information only redward of 1260\AA\ rest-frame to ensure unbiased reconstruction. We constructed two samples of 30 QSOs, from X-Shooter and SDSS both spanning $3.5<z<4.5$. Importantly, the purpose of this comparison study was not to champion a single, best performing reconstruction pipeline but rather to explore the relative performance of these pipelines over a range of QSOs with broad observational characteristics to infer general trends. In summary, we find machine learning approaches in general provide the strongest ``best guesses" but underestimate the accompanying statistical uncertainty, although these can be recalibrated, whilst pipelines that decompose the spectral information, for example principal component or factor analysis generally perform better at predicting the \lya{} profile. Further, we found that reconstruction pipelines trained on SDSS QSOs performed similarly on average for both the X-Shooter and SDSS samples indicating no discernible biases owing to differences in the observational characteristics of the training set or QSO being reconstructed, although the recovered distributions of reconstructions for X-Shooter were broader likely due to an increased fraction of outliers.
\end{abstract} 
\begin{keywords}
quasars: emission lines -- quasars: general -- cosmology: observations -- cosmology: theory
\end{keywords}

\section{Introduction}

Interpreting observed quasar (QSO) spectra plays an important role in our understanding of the cosmos. For example: (i) studying the emission line properties can reveal insights into the central supermassive black hole (SMBH) and the internal structure of the SMBH powered accretion disk \citep[e.g.][]{Blandford:1982,Peterson:1993,Kaspi:2000,Peterson:2004,Vestergaard:2006} or (ii) as sight-line probes through the intergalactic medium (IGM) to explore the baryon content, ionisation state or its metallicity (for a more detailed summary see recent reviews by e.g. \citealt{Becker:2015b,McQuinn:2016,Fan:2023}). For this work, we focus on the latter case, in particular the use of QSOs for exploring the epoch of reionisation (EoR), the baryonic phase transition from an initially neutral IGM following the surface of last scattering to its almost complete ionisation by the cumulative ultra-violet (UV) ionisations of the first stars, galaxies and active galactic nuclei. 

QSOs are intrinsically bright and typically exhibit prominent Lyman-alpha (\lya{}) emission powered by thermal emission from the accretion disk. At $z\lesssim5$, owing to the strength of the \lya{} scattering cross section, diffuse clouds of neutral hydrogen scatter the \lya{} photons away from the observer giving rise to distinct absorption features in the spectra blueward of the \lya{} emission of the QSO. As QSO sight-lines probe vast cosmic distances through the IGM, they intersect numerous neutral hydrogen absorbers producing a series of discrete narrow absorption features, known as the \lya{} forest \citep{Rauch:1998}, which can be directly interrogated to divulge the properties of the intervening IGM (e.g. ionisation state and metallicity).

Unfortunately, by $z\gtrsim5$, the strength of resonant absorption by even trace amounts of residual neutral hydrogen ($x_{\hi} \gtrsim 10^{-4} - 10^{-5}$) in the IGM saturates the transmission of \lya{} photons \citep[e.g.][]{Gunn:1965,Fan:2006p4005}. Order unity fluctuations in the neutral fraction due to reionisation are indistinguishable from fluctuations in the ultraviolet background, density or temperature \citep[e.g.][]{D'Aloisio:2015,Keating:2018}. This renders direct studies of individual \lya{} absorption features and their connection to the IGM during the EoR essentially impossible. However, large-scale fluctuations in opacity can yield insights \citep[e.g.][]{Bosman:2018,Bosman:2022}.

Importantly, the scattering profile of \lya{} photons is Lorentzian, with a suppression of roughly 5-6 orders of magnitude in its amplitude at fairly modest velocity offsets away from the resonant core. Thus, as the IGM becomes increasingly neutral, absorption by these wings can become sufficiently large to imprint a smooth absorption feature onto the intrinsic QSO spectrum. This IGM damping wing absorption \citep{Rybicki1979,MiraldaEscude:1998p1041} can then in principle be used to provide constraints on the IGM neutral fraction during the EoR if we can infer it from the observed QSO spectrum. Doing so quantitatively requires the ability to statistically predict the intrinsic (unattenuated) \lya{} emission line from the observed QSO spectrum.

Within the literature several approaches have been developed to predict the intrinsic \lya{} emission line of high-$z$ QSOs \citep[][Meyer et al., in prep]{Greig:2017a,Davies:2018a,Dominika:2020,Fathivavsari:2020,Reiman:2020,Bosman:2021,Liu:2021,Sun:2022}. Albeit different in their methodology, fundamentally these are built on the same core expectation that the intrinsic properties of the \lya{} emission line are strongly correlated with other observable emission lines \citep[e.g.][]{Boroson:1992p4641,Shang:2007p4862,Kramer:2009p920} or spectral information sufficiently redward of \lya{} to be unaffected by the smooth attenuation of a neutral IGM. Thus, given a sufficiently large training sample of unattenuated QSOs (i.e. $z\lesssim5$) a predictive model can be developed to connect the unattenuated (observable) red-side information to reconstruct the intrinsic \lya{} profile.

Coupled to either an analytic model or simulations of the IGM damping wing, several of these \lya{} reconstruction pipelines have been independently applied to four of the eight\footnote{For a continually updated list of $z>5.7$ QSOs see \citet{Bosman:2020list}} known $z\gtrsim7$ QSO in the literature suitable for such an analysis\footnote{Of the remaining four,  HSCJ2356+0017 \citep[$z=7.01$,][]{Matsuoka:2019b} and HSCJ1243+0100 \citep[$z=7.07$,][]{Matsuoka:2019a} are too faint while both J0038-1527 \citep[$z=7.02$,][]{Wang:2018} and the current highest redshift QSO to date, J0313-1806 \citep[$z=7.64$,][]{Wang:2021}, are broad absorption line QSOs.}: ULASJ1120+0641 \citep[$z=7.08$][]{Mortlock:2011p1049}, ULAS1342+0928 \citep[$z=7.54$,][]{Banados:2018}, DESJ0252-0503 \citep[$z=7.00$,][]{Yang:2019} and J1007+2115 \citep[$z=7.51$,][]{Yang:2020} to yield the first direct constraints of ongoing reionisation \citep[][]{Greig:2017,Banados:2018,Davies:2018,Greig:2019,Reiman:2020,Dominika:2020,Wang:2020,Yang:2020,Greig:2022,Dominika:2024}. Qualitatively speaking, despite their inherently different methodologies, the resultant constraints on the IGM neutral fraction during the EoR are broadly consistent, generally driven by the relatively large modelling uncertainties.

Crucially, despite the numerous available \lya{} reconstruction methods available in the literature, almost no direct comparison of the \lya{} profile predictions for the same objects have been performed\footnote{Note, \citet{Bosman:2021} explored several methods in the literature to predict the placement of the \lya{} forest continuum blue-ward of the QSO proximity zone.} (although see e.g. \citealt{Davies:2018,Greig:2019,Greig:2022} for some qualitative discussions). In this work, we aim to rectify this by directly comparing the predicted \lya{} profiles from the aforementioned reconstruction methods on a unified dataset of observed QSO spectra. In particular, we perform this study at $z\lesssim5$ in order to be able to directly compare against the intrinsic (almost unattenuated) \lya{} profile. Further, we consider two distinct samples of QSOs selected from X-Shooter \citep{Vernet:2011} and the Baryon Oscillation Spectroscopic Survey (BOSS; \citealt{Dawson:2013p5160}) observed with Sloan Digital Sky Survey (SDSS) telescope \citep{Gunn:2006,Smee:2013}. This particular distinction was designed to explore the applicability of these reconstruction methods to determine if the quality of the training set or the target QSO to be reconstructed can bias the predictions. For example, almost all reconstruction methods are trained on lower resolution BOSS spectra but are applied to $z\gtrsim7$ QSOs obtained from different instruments and observational characteristics (typically X-Shooter \citealt{Vernet:2011}).

Finally, and most importantly, to prevent biasing the results of this reconstruction comparison, the provided QSO datasets for each \lya{} pipeline author were transformed to rest-frame, de-identified and blinded below 1260\AA. Thus, when predicting the \lya{} line profiles, there was no prior knowledge of the correct (intrinsic) one. This step ensured each author ran their own pipeline, rather than having one individual learn to run/re-train the various methods and potentially apply them incorrectly. Note, the purpose of this study is to compare how well each individually trained method performs at reconstructing the same set of objects given their own unique assumptions on the size and quality of the training set. That is, how universal and robust are the methods at reconstructing the same broad range of objects. If instead we were interested in performing a rigorous comparison of the actual reconstruction pipeline methodology, each pipeline would need to be re-trained on the exact same training set. This would be a significant undertaking and as such we defer it to future work.

We structure this investigation as follows. In Section~\ref{sec:setup} we introduce the blind QSO datasets, summarise the various \lya{} reconstruction pipelines and provide any other general discussions. In Section~\ref{sec:qualitative} we provide some individual blind QSO reconstructions from both samples along with some associated discussions. Next, in Section~\ref{sec:quantitative} we perform a more quantitative comparison of individual and sample averaged properties before providing some over arching discussions and our conclusions in Section~\ref{sec:conclusions}.

\section{Methodology} \label{sec:setup}

\subsection{Data sample}

One of the common assumptions made when reconstructing \lya{} profiles is that these pipelines can be applied universally to QSOs observed by any instrument. Typically, to obtain a sufficiently large training set of QSOs, almost all pipelines are based on BOSS QSO datasets. However, can a BOSS trained pipeline be used to reconstruct a QSO of different quality (i.e. S/N and resolution)? We explore this by constructing two independent samples of 30 QSOs which differ by these quality metrics. This choice in sample size is fairly arbitrary, but sufficient to be able to provide a statistically robust exploration.

For our first sample, Sample 1, we select 30 QSOs observed by X-Shooter \citep{Vernet:2011}. In particular, these are randomly drawn from the XQ-100 sample \citep{Lopez:2016}, which contains 100 X-Shooter spectra spanning $3.5 < z < 4.5$, with a resolving power of $R\sim5400-8900$ and a median S/N $\sim30$. This constitutes our high-quality QSO sample, with which we can infer if any biases are present owing to the differences in QSO spectra quality and/or from using a different instrument than the training set data.

The second sample of 30 QSOs (Sample 2) comes from BOSS, in particular the DR16Q catalogue \citep{Lyke:2020}. BOSS has a typical resolving power of $R\sim2000$ and we randomly select QSOs between $3.5 < z < 4.5$ (consistent with Sample 1 above) with a S/N of $\sim10-30$. These QSOs are equivalent to the quality of QSOs used in most training sets, and thus serve as a baseline for QSO pipeline performance. That is, at the very least the reconstruction pipelines should perform well on spectra of similar observational characteristics.

Note, the choice of selecting QSOs at $3.5 < z < 4.5$ is three-fold. Firstly, at these redshifts there should be no significant attenuation from neutral gas redward of \lya{} in the IGM (outside of the \lya{} forest), in which case the observed \lya{} emission line is the intrinsic \lya{} line profile (modulo a small number of intervening absorption). Next, the reconstruction pipelines assume that there is no significant redshift evolution in the intrinsic \lya{} line profile. That is, we can construct our training sets at low-$z$ where there is a wealth of data and apply these methods to any higher redshift QSO. Pipeline training sets are typically dominated by QSOs at $2 < z < 2.5$, thus our sample at $3.5 < z < 4.5$ allows us to explore any potential impact due to redshift evolution (see also Appendix~\ref{sec:z_evolution}). Finally, the vast majority of pipelines draw their training sets at $z\lesssim2.5$ from BOSS. Thus, by selecting at higher redshifts, we generally avoid the risk of these blind QSOs being within the training sets and thus potentially biasing our results towards being overconfident in their performance (i.e. easier to recover the correct QSO profile).

Although our blinded QSOs were selected at random, they were visually checked to ensure each QSO did not contain substantial \lya{} absorption features which would otherwise make it difficult to determine the intrinsic spectrum.   Further, these QSOs were selected to ensure there were no broad absorption line features or proximate damped \lya{} absorbers along the line of sight. Other than this, no further visual identification or classification was performed, thus it is possible that some of these QSOs might contain spurious features or characteristics that would otherwise make them unsuitable for science grade studies. However, since the goal of this work is to perform a comparison rather than to derive a scientific result, there is no harm in including such potential contaminants.

\subsection{Blind Sample Setup}

Following the identification of our two QSO datasets, we construct the blinded data that was provided to each of the individual \lya{} reconstruction pipeline authors. In total, each participant was provided with 60 rest-frame QSO spectra, blinded blueward of 1260~\AA\ and de-identified (i.e. provided an arbitrary name). These were separated into two samples (Sample 1 and 2 as described above), however at the time the authors had no knowledge of the instrument from which these spectra were obtained. Only once all results were received from all reconstruction pipelines were the details of the individual QSOs and the true \lya{} line profile provided to the authors.

For Sample 1 (X-Shooter), the participants were provided QSOs with spectral coverage spanning [1260, 3000]~\AA\ and the spectra were re-binned onto 0.2 \AA\ bins. This binning was chosen to obtain a similar rest-frame sampling rate to the BOSS spectra. Several reconstruction pipelines perform their own re-processing of the spectra including re-binning (coarser) thus the choices adopted here will not impact the performance of the reconstructions. The chosen spectral coverage was adopted to include the \mgii{} line ($\lambda=2800.26$\AA), which is used by a handful of the reconstruction methods, generally speaking the principal component analysis (PCA) pipelines (see Section~\ref{sec:pipelines}).

For Sample 2 (SDSS), participants were provided with spectral coverage spanning [1260, 2100]~\AA\ at the default BOSS wavelength sampling (i.e. no re-binning performed). The reduced spectral coverage owes to the fact that the \mgii{} lines falls out of the BOSS spectral window ($\lambda=4000-10000$\AA) for our $3.5 < z < 4.5$ selected QSOs. 

\subsection{QSO reconstruction pipelines} \label{sec:pipelines}

Having outlined the blinded QSO data above, we now introduce the individual \lya{} profile reconstruction pipelines involved in this comparison study. At the time of this study, all known \lya{} reconstruction pipelines were included into this analysis. Below we outline each of the participating pipelines, providing a summary of the main assumptions, methodology and any other information relevant for this comparison study. Further, we also provide a brief summary of this information in Table~\ref{tab:pipelines}. For more detailed information on any of the pipelines, we defer the reader to the introductory publications mentioned below.

\begin{table*}
%\tiny
\caption{A brief summary of the various QSO reconstruction pipelines used within this work (see text for further details).}
%\begin{adjustwidth}{-0.6cm}{}
%\begin{adjustwidth}{}{}
%\scalebox{0.8}[0.75]{
%\begin{adjustwidth}{-2cm}{}
\begin{tabular}{@{}lcccccc}
\hline
Pipeline & Methodology for QSO reconstruction & \# of QSOs in training set & S/N & $z$-coverage & Catalogue \\
\hline
Bosman & PCA decomposition; predict 10 blue-side components & 14,029 & > 10 & (2.25, 3.5) & BOSS DR14Q \\
\vspace{0.8mm}
 & based on 15 red-side components &  &  &  \\
(ftof) & Same as above with a slightly modified pre-processing step \\
\hline
Davies & PCA decomposition; predict 6 blue-side components  & 12,764 & >7 & (2.09,2.51) & BOSS DR12Q \\
 & based on 10 red-side components &  &  &  \\
 \hline
 \v{D}urov\v{c}\'{i}kov\'{a} & PCA decomposition; predict 36 blue-side components based on  & 13,703 & > 7 & (2.09,2.51) & BOSS DR14Q  \\
 (QSANNdRA) & 63 red-side components using an ensemble of artificial neural networks & & & \\
 \hline
 Greig  & Covariance matrix reconstruction using emission lines  & 30,166 & >6.5 & (2.0, 3.5) & BOSS DR16Q \\
  & modelled as Gaussian profiles & & & & \\
\hline
Fathivavsari & Neural network based reconstruction of QSO flux per pixel & 17,870 & > 15 & (2.0,4.3) & BOSS DR14Q \\
\hline
Liu (iQNet) & Auto-encoder neural network architecture to & 63 & > 5 & < 1.0 & HST (COS) \\
\vspace{0.8mm}
(HST) & predict the blueward QSO continuum & (extended to 12,000 mock) &  &  & \\
(HST + SDSS) & Same method but combining mock HST spectra and BOSS QSOs & 3196 & > 5 & (2.0, 2.5) & BOSS DR16 \\
\hline
 Meyer & Predict using an information maximising variational auto encoder & 9,926 & > 7 & (2.09,2.51) & BOSS DR16Q \\
   & (final training set artificially enlarged by a factor of 10) & & & & \\
 \hline
 Reiman  & PCA decomposition; posterior predictions achieved  & 13,703 & > 7 & (2.09,2.51) & BOSS DR14Q \\
 (Spectre) & using normalising flows & & & & \\
 \hline
 Sun (QFA) & Latent factor analysis combined with unsupervised learning & 50,000 & > 2 & (2.0,3.5) & BOSS DR16Q \\
\vspace{0.8mm}
 & to simultaneously learn the QSO continuum and \lya{} forest & & & & \\
\vspace{0.8mm}
 (Analogue) & Same as QFA but trained per-object with 500 nearest neighbours & 500 &  &  &\\
(Analogue wNN) & Same as QFA Analogue but with an additional neural network & 500 & &  \\ 
 & reconstruction of the latent factor amplitudes & & &  \\ 
\hline
\end{tabular}
%}
\label{tab:pipelines}
%\end{adjustwidth}
\end{table*}

\subsubsection{Greig}

Taking direct advantage of the known correlations between the \lya{} emission line and other high-ionisation emission lines, in \citet{Greig:2017a} a covariance matrix based method to predict the \lya{} profile was developed. This assumes individual emission lines can be fit by either a single or double component Gaussian profile, characterised by the line width, amplitude and velocity offset from systemic. For the \lya{} and \civ{} emission line profile a double component Gaussian profile is preferred while for all other lines a single Gaussian profile was deemed sufficient.

The original covariance matrix was determined from 1673 visually selected QSOs from DR12 \citep{Alam:2015p5162} spanning $2.0 < z < 2.5$ with S/N $>15$. Each observed QSO spectrum was fit using Monte-Carlo Markov-Chain (MCMC) including all known emission lines, a simple power-law continuum (amplitude and spectral index) and a variable number of Gaussian profiles to characterise any absorption features. Following this, the covariance matrix is determined for just the emission line parameters; those describing the \lya{} profile and the other prominent, high-ionisation emission lines (e.g. \siv{}, \civ{} and \ciii{}). Importantly, for this work, we use a new covariance matrix (Greig et al., in prep) which additionally includes the \nv{} emission line and is based off the BOSS DR16Q catalogue \citep{Lyke:2020} using 30,166 QSOs with S/N $> 6.5$ and spanning $2.0 < z < 3.5$.

\lya{} line reconstruction is then performed by adopting a $N$-dimensional normal distribution describing the emission line parameters with the above covariance matrix. We MCMC fit the QSO to be reconstructed (using only spectral information redward of 1260~\AA), simply extrapolating the recovered two parameter power-law continuum and add to this our $N$-dimensional model using the information from fitting to the \siv{}, \civ{} and \ciii{} lines (the recovered maximum $a$-posteriori values for these emission line parameters). This results in a 9-dimensional normal distribution which provides a fully probabilistic determination of the \lya{} + \nv{} line profile on top of the two parameter power-law continuum.

Note, in the past when applying this \lya{} reconstruction method an additional prior on the QSO flux near $\sim1250$\AA\ was applied to ensure the reconstructed profiles matched the amplitude of the observed spectrum. This had the benefit of reducing the relative amplitude of the modelling uncertainties by using additional information from the observed spectrum. However, in this work, we cannot apply this prior as we do not have spectral information blueward of 1260~\AA. As a result, the model uncertainties will be notably larger than previous implementations of this pipeline.

\subsubsection{Davies}

Rather than modelling each emission line as a Gaussian profile with three parameters, a more optimal method is to compress the wealth of spectral information into a small, finite number of components using PCA decomposition \citep{Boroson:1992p4641,Francis:1992p5021,Suzuki:2005,Paris:2011p4774}. This nonparametric approach was utilised to predict the \lya{} profile by \citet{Davies:2018a}. Here, the blueward component of the QSO spectrum ($<1280$~\AA) can be predicted based on the redward information [1280, 2900]~\AA. To extract the PCA components the QSO spectra were first fit with an automated continuum-fitting method introduced in \citet{DallAglio:2008} based off the earlier work of \citet{Young:1979} and \citet{Carswell:1982}.

In total, this model decomposes QSO spectra into 6 blue-side components and 10 red-side components, whereby these blue-side components are predicted using a projection matrix based off the red-side information. To construct this projection matrix, a training set of 12,764 BOSS QSOs were obtained from the DR12 quasar catalogue \citep{Paris:2017} spanning $2.09 < z < 2.51$ and S/N $>7$.

Note for this work, this PCA pipeline was retrained for Sample 2 (SDSS) as these blind QSOs contain reduced redward spectral information [1280, 2100]~\AA. The original method is used for Sample 1 (X-Shooter) which includes the \mgii{} emission line.

\subsubsection{QSANNdRA - \v{D}urov\v{c}\'{i}kov\'{a}}

Extending this approach of PCA decomposition, \citet{Dominika:2020} connected the blue and red-side PCA components with an artificial neural network (ANN). Within this approach, QSANNdRA, the number of PCA components were significantly expanded out to 36 on the blue and 63 on the red-side, respectively. This expansion of the PCA components was designed to pick up 99 per cent of the variance in the QSO training set.

These red and blue-side components were connected by a four layered fully connected feed forward network, with the input layer corresponding to the red-side PCA components and the output the blue-side predictions. QSANNdRA was trained on a slightly larger training set, 13,703 QSOs obtained from the BOSS DR14 quasar catalogue \citep{Paris:2018} with the same redshift and S/N cuts as \citet{Davies:2018a}; $2.09 < z < 2.51$ and S/N $>7$. The QSO spectra within this training set are smoothed using a publicly available custom routine (\url{github.com/DominikaDu/QSmooth}) which includes a random forest to exclude DLA QSOs and identify intervening absorbers before extracting the PCA components. For this work, QSANNdRA was re-trained to make use of the slightly extended spectral coverage of the blind QSO data ($>1260$~\AA) compared to its original coverage ($>1290$\AA).

In an attempt to circumvent the issue of ANNs providing a prediction with no corresponding uncertainty, QSANNdRA was independently trained 100 times (i.e. bootstrap aggregating). Note however, this statistical uncertainty accounts for variations in the network training and not the inherent scatter in the QSO population. Thus, the associated errors from QSANNdRA (typically smaller than all other methods) have a different statistical meaning to the uncertainties from other methods, which will be important for later discussions.

\subsubsection{Fathivavsari}

Instead of decomposing the spectral information into PCA components, \citet{Fathivavsari:2020} developed a neural network approach to predict the QSO flux in each spectral bin, rather than relying on some implicit shape (Gaussian profiles or PCA components). The QSO flux is predicted between [1170, 1290]~\AA\ (encompassing \lya{} and \nv{}) based directly off the observed flux within wavelength bins between [1370, 1430]~\AA\ (\siv{}), [1505, 1590]~\AA\ (\civ{}) and [1865,1945]~\AA\ (\ciii{}). To extract the relevant spectral information the observed spectra are fit using a custom automated pipeline which iteratively applies Savitzky-Golay filtering \citep{Savitzky:1964} and removes pixels who deviate by more than two standard deviations from the fit to the spectrum.

To predict the QSO flux, the trained network consists of an input layer containing the total number of wavelength bins describing the redward spectral information (450 neurons) and an output layer consistent with the number of wavelength bins for the blueward information to be predicted (250 neurons). These two layers are then connected by three hidden layers which are fully connected, feedforward with error back propagation.

This approach is trained on 17,870 BOSS QSOs from the DR14 quasar catalogue \citep{Paris:2018}, spanning a broader redshift range ($2.0 < z < 4.3$) but S/N$>15$. Note, this expanded redshift range means that it is possible that some of the blind QSOs from Sample 2 (SDSS) could appear in the training set. However, the potential appearance of these objects should not bias the results in any meaningful way as the trained network is not tied directly to individual objects in the sample (i.e. it is based on the distribution as a function of wavelength).

\subsubsection{Spectre - Reiman}

Extending on the machine learning - PCA approach of QSANNdRA, \citet{Reiman:2020} developed Spectre, which casts the prediction of the blue-side PCA components from the red-side information as a conditional density estimation problem. This enables Spectre to predict plausible blue-side spectral information drawing from a full probabilistic distribution. Spectre was trained on the same dataset as \citet{Dominika:2020}, consisting of 13,703 QSOs obtained from the BOSS DR14 quasar catalogue \citep{Paris:2018} at $2.09 < z < 2.51$ and S/N $>7$.

Spectre utilises normalising flows \citep[see e.g][for a review]{Papamakarios:2019}. First, input QSO spectra are processed following the same pipeline as \citet{Dominika:2020}, before performing PCA decomposition. However, instead of placing the PCA components as direct inputs, an encoder network is used to extract relevant information from the redward spectral information which is connected to a four layered fully connected network. By transforming random Gaussian samples through a series of ten coupling transforms, each conditioned on the redside information, the distribution of redward information can be transformed into a probabilistic distribution describing the blueside information. From this, plausible blueward QSO flux profiles can be drawn.

Importantly, for this work we utilise the publicly available implementation of Spectre (\url{github.com/davidreiman/spectre}). We retrained Spectre using the updated data provided as improvements to QSANNdRA as noted above, and verified this produced similar results to the original provided examples. Unfortunately, the main authors with domain knowledge of Spectre had left academia at the time of this study and were unavailable. Therefore, we caution about over interpreting the results based on Spectre, as we are assuming our implementation of Spectre is suitable for this blind comparison study. Quite possibly we have overlooked crucial steps which may have been detrimental to the performance of our own implementation of Spectre.

\subsubsection{iQNet - Liu}

An alternative deep learning approach, the intelligent Quasar continuum neural network (iQNet), was developed by \citet{Liu:2021}. The original methodology for iQNet differs slightly to the aforementioned approaches as it was designed specifically for $z\lesssim5$ QSOs (no IGM attenuation) and subsequently to take as input the spectral information at [1216, 1600]~\AA\ to output information at [1020, 1600]~\AA, focussing more on the \lya{} forest. However, for this comparison, iQNet was modified instead to take as input the spectral information at [1260, 1600]~\AA\ and then predict the QSO flux between [1180, 1260]~\AA.

iQNet treats the input rest-frame QSO as a 1 dimensional array, extracting relevant spectral information using a simple auto-encoder architecture. Again the input and output layers are designed to be the redward and blueward spectral information, fully connected by seven additional neural layers which perform the encoding and decoding. The redward QSO information, [1260, 1600]~\AA, is treated as the sum of the intrinsic QSO continuum, random noise and intervening absorption line systems imprinted on the spectra. The network architecture then returns an estimate of the full intrinsic QSO continuum between [1180, 1260]~\AA. 

For this work, two separate networks were trained. First, using the original QSO training set obtained from the Hubble Spectroscopic Legacy Archive Data Release 2 \citep{Peeples:2017} observed with the Hubble Space Telescope (HST). Here, 63 QSOs at $z<1$ with median S/N $>5$ per resolution element were selected to train iQNet (HST). To increase this limited dataset, a Gaussian mixture model was implemented on the PCA components of these HST QSOs to a training set of $12,000$ mock plus real HST spectra. The second network, iQNet (HST+SDSS), includes a joint training set including the previous training set for the HST network along with BOSS DR16 QSOs \citep{Ahumada:2020}. For the BOSS component, 3196 QSOs were selected following visual inspection with a median S/N $>5$ per pixel and between $2.0 < z < 2.5$ to limit the contamination of the \lya{} forest on the QSO continua.

\subsubsection{Bosman}

As mentioned earlier, \citet{Bosman:2021} performed a comparison study of available reconstruction techniques for predicting the QSO continuum for the \lya{} and \lyb{} forest ($\lambda < 1210$\AA). In doing so, updated versions of the \citet{Davies:2018a} PCA-based approach were developed which were then further extended in \citet{Chen:2022}.

Specifically, two networks were trained, which differed in the pre-processing step of the training data. The first, (Bosman) extracts the PCA components directly from the observed spectrum. For the second (Bosman-ftof), the observed spectrum is first fit by a slowly varying spline (to better handle missing spectral information) which is then PCA decomposed. Both of these use 15 and 10 PCA components for the red and blue side, respectively, an increase in the number of components from the original \citet{Davies:2018a} approach. These pipelines were trained on a sample of 14029 QSOs, which was cleaned by hand to remove all proximate DLAs, extracted from the DR14 quasar catalogue \citep{Paris:2018}, with a redshift spanning $2.25 < z < 3.5$ and a S/N $>10$. Ultimately, these two approaches should perform similarly, although some slight differences might occur for blind QSOs with missing spectral information redward of 1280\AA.

\subsubsection{Meyer}

In Meyer et al. (in prep.) an information maximising variational auto encoder (InfoVAE, e.g. \citealt[][]{Zhao:2017}) is adopted to predict the \lya{} line profile. Here, the InfoVAE is trained as a denoising auto encoder where corrupted 1D images of QSO spectra are input and then reconstructed to predict the QSO continua between [1180, 1260]~\AA. For this, learning is applied to train neural networks to perform the compression (encoding) and reconstruction (decoding) based on the spectral properties of the QSOs.

To train this method, QSOs were extracted form the BOSS 16th data release QSO catalogue \citep{Lyke:2020} with a median S/N $>7$ and spanning $2.09 \lesssim z \lesssim 2.51$. The SDSS spectra are smoothed following the approach initially developed by \citet{DallAglio:2008} and refined by \citet{Bosman:2021}. Following this smoothing and removing QSOs rejected on the basis of being poor fits or containing DLAs, BALs or strong absorption systems, 9926 QSOs are retained for training. To improve the robustness of the method the training sample is increased by resampling each QSO 10 times by randomly adding an associated velocity shift to the original redshift (-2000,2000)~km~s$^{-1}$, applying random masks mimicking the treatment of sky lines in near-infrared quasar spectroscopy, additional noise is added to the QSO flux based on the noise properties of SDSS before finally each spectrum being smoothed again following the \citet{DallAglio:2008} method.

\subsubsection{QFA - Sun}

Extending on the idea of PCA decomposition, \citet{Sun:2022} introduced an unsupervised and probabilistic approach to predict the QSO continuum, named QFA. The principal difference of QFA is the unsupervised aspect, where it is designed to use the entire spectrum to infer the posterior of the intrinsic profile. This differs from all the other methods which perform supervised learning to use the estimated smoothed QSO continua to predicted the mapping of the redward information to the blueward information. To characterise the QSO continuum, QFA employs latent factor analysis \citep[e.g.][]{Bartholomew:2011,Loehlin:2017} which is a powerful statistical method which assumes high-dimensional correlated data can be expressed as a set of lower-dimensional latent factors. Effectively, it is a more flexible version of PCA decomposition. In total, QFA adopts 8 latent factors to describe the QSO spectral information.

Originally, QFA was trained over [1030, 1600]~\AA, with the primary aim of estimating the QSO continuum for the \lya{} forest, which is achieved by a novel method to jointly model the QSO continuum (with latent factors) and a normal distribution modelling \lya{} forest absorption. For this work, QFA was re-trained to predict the blueward information, [1030, 1260]~\AA, of the \lya{} forest and \lya{} emission region from available redward information at [1260, 2050]~\AA.

QFA builds a high dimensional multivariate Gaussian model (i.e. the shape of the latent factors, \lya{} forest model, plus other nuisance parameters) that can fully describe the properties of the entire QSO distribution. This model is then trained, given an input training set to recover the maximum likelihood model (of continuum plus \lya{} forest absorption) that best describes the data. The output, amongst other quantities are the latent factors, which span the full spectral range, [1030, 2050]~\AA, that best characterise the QSO spectra. For this work, QFA was retrained on a random selection of 50,000 QSOs from BOSS DR16Q \citep{Lyke:2020} spanning $2.0 < z < 3.5$ and S/N $> 2$. Note that for QFA it is recommended to rescale the recovered model uncertainties by a factor of three, which we perform throughout this work.

Importantly for this comparison, QFA does not distinguish between the red/blue spectral information since it is trained to obtain latent factors spanning the full spectral range, [1030, 2050]~\AA. This differs from most of the approaches outlined earlier, where the training data is fit independently with respect to the red and blueward spectral information before using the correlations between the red/blue spectral information to predict the blueside components given the fit redward components (e.g. projection matrix or ANN in the case of PCA). Therefore, when QFA is applied to reconstruct the intrinsic QSO profile, the QFA model is fit to just the redward spectral information ($>1260$\AA) which determines the posterior distribution of latent factor amplitudes that best characterise the available redward information. Sampling these recovered posteriors of the amplitudes multiplied by the latent factors spanning [1030, 2050]~\AA\ then yields the full intrinsic QSO profile.

\subsubsection{QFA - analogue}

The idea of QFA is to train a model consisting of a large number of free parameters that can fully describe the entire QSO population (QSO continuum, \lya{} forest absorption plus other nuisance parameters). However, in practice this effectively results in learning the latent factors that describe the median QSO population and it assumes that all QSOs can be adequately characterised by these latent factors. If the QSO to be reconstructed exhibited spectral properties that differed to these underlying latent factors, it would not be able to provide a reasonable prediction of the blueward information.

Therefore, an alternative version of the QFA model was developed, which instead relies on training the same model as described above, except using a training set of 500 analogue QSOs that exhibit similar redward properties as the target QSO. To identify these objects, a nearest neighbour search is performed over the provided redward spectral region, [1260, 2050]~\AA, selecting the 500 QSOs with the smallest euclidean distance (Equation~\ref{eq:NN}). Following the identification of the nearest neighbours QFA is retrained for each individual blind QSO.

\subsubsection{QFA - analogue with NN}

QFA does not distinguish between the blueward and redward spectral information. To explore whether this has an impact on the performance of the QSO reconstructions we considered a third QFA model which includes a form of blueward conditioning given the available redward spectral information. Following a similar approach as above, QFA is first trained on the same 500 nearest neighbour analogues. However, instead of simply drawing from the posterior distribution of latent factor amplitudes that best describe the redward information which equally yields the blueward information owing to the spectral coverage of the latent factors, [1030, 2050]~\AA, the blueward information is instead directly predicted based off the redward information.

Specifically, for each of the 500 QSOs in an individual training set, the amplitudes of the latent factors are determined from fitting over the entire spectral range, [1030, 2050]~\AA, and from only the redward information, [1260, 2050]~\AA. Then, given this data, a neural network is trained to correct the latent factor amplitudes obtained from the redward information to be consistent with the distribution obtained from the full spectral coverage. 

\subsection{Discussions} \label{sec:init-disc}

In total, we have 13 available models to predict the \lya{} emission line flux of our blind QSOs. These span a broad range of methodologies and assumptions enabling a detailed exploration of reconstruction performance. Here, we summarise the main observations and discussion points in preparation for the comparison.

Firstly, only the iQNet (HST) model is trained on a different dataset than BOSS QSOs. In particular it is trained on HST spectra that are re-sampled onto 0.05\AA\ bins, which contrasts to the nominal BOSS wavelength sampling of $\sim0.2$\AA. With a training set of significantly higher resolution, the observed QSO spectra will differ in their observational characteristics as they are convolved by a narrower point spread function (PSF). As a result, the spectral features will be sharper (i.e. narrower and more sharply peaked) than the same QSO observed by BOSS. The impact of this will be interesting as it explores the impact of the quality of the QSO training set on the relative performance of the \lya{} profile reconstructions. However, it will be difficult to completely disentangle this from the fact that this sample is also drawn from notably lower redshifts ($z<1$). Therefore it is possible that the properties extracted from the HST QSOs used for their work may differ from the properties of the blinded QSOs at $z\sim3.5-4.5$.

All reconstruction methods assume that any correlations between the various emission lines and/or any other extracted trends (e.g. between the PCA components) do not change with either redshift or luminosity. That is, we can apply our reconstruction pipelines on a QSO at any redshift (i.e. $z\gtrsim6$) or luminosity irrespective of the characteristics of the data selected for the training set. The benefit of this is that we can access the significantly larger samples of analogue low-$z$ QSOs for building the training sets of our reconstruction methods. This is despite some loose evidence for modest redshift evolution out to $z\sim5-6$ based off QSO composite spectra \citep{Becker:2013}, albeit with a small number of objects, or the known correlations between line equivalent widths and QSO luminosity \citep{Dietrich:2002} (see also \citealt{Sun:2022}). Approaching $z\sim7$ however, evidence points to a more rapid evolution in the QSO features as evidenced by the notable change in \civ{} blueshifts \citep{Meyer:2019} or properties of the broad absorption line QSOs \citep{Bischetti:2022,Bischetti:2023}. However, at these redshifts we are significantly restricted by the number of available QSOs.

To explore these assumptions, we split the BOSS DR16Q sample into narrow redshift or luminosity bins to search for indicators of either redshift evolution (Appendix~\ref{sec:z_evolution}) or QSO luminosity (Appendix~\ref{sec:L_evolution}) on reconstructions of the blind QSO sample. For both, we recover faint hints for evolution in the predicted \lya{} profiles, but the relative amplitude in evolution is well within the statistical uncertainties and thus equally consistent with no evolution. However, we do note hints for stronger evolution due to luminosity for QSOs of increasing equivalent width (as seen by \citealt{Dietrich:2002}), particularly around the amplitude of the \lya{} line. However, the other redward emission lines used for reconstructing the QSOs tend to also correlate similarly with luminosity as the \lya{} line therefore it is unlikely to have a strong effect on the \lya{} reconstructions. Nevertheless, care may need to be taken when constructing the training samples to ensure they are similar to the properties of the QSOs to be reconstructed. Although redshift evolution does not appear to be prevalent, we note that several of the reconstruction pipelines, mostly the PCA based approaches, have a relatively narrow redshift coverage for their training set of $2.0 < z < 2.5$ which differs from the $3.5 < z < 4.5$ redshift coverage of the blind QSO data. Although this difference in redshift is relatively small, $\Delta z \sim 1.5$, if significant redshift evolution existed, and impacted the reconstructions, this should result in a systematic bias in the predicted \lya{} profiles from these pipelines. Therefore, these results will provide another test of redshift evolution. Note however, that these tests of redshift evolution are only valid for the redshifts considered in this work (i.e. $z\lesssim5$). We cannot test for the more significant evolution observed at $z\sim7$ owing to the relatively low number of objects and the fact that these objects are likely strongly attenuated by a significantly neutral IGM in which case we have no ability to $a$-priori estimate the true intrinsic \lya{} profile.

Finally, for this blind QSO comparison project we need to estimate the truth intrinsic profile (ground truth) of the original QSOs. To estimate this, the \citet{Greig:2017a} fitting pipeline was applied to the original QSO spectra. This pipeline performs a maximum likelihood fit simultaneously varying a power-law continuum, Gaussian emission lines profiles and can identify and fit any number of \lya{} absorbers with a Gaussian profile. The benefit of this is that it can automate the estimation of the QSO continuum around \lya{}, which is the primary purpose of this comparison. The fits to the intrinsic profile were then visually inspected to ensure validity.

Although the \citet{Greig:2017a} methodology is used for estimating the intrinsic truth, it should not bias the results of the associated \lya{} reconstructions obtained using the same pipeline. The reconstructions are based off the covariance matrix of line correlations determined from the entire QSO training set (i.e. statistical distribution of the entire sample) and only uses information redward of 1260\AA, while the MCMC pipeline fits the full wavelength coverage, [1180, 2050]~\AA, of the individual object (i.e it does not rely on the line correlations). Note also, \citet{Bosman:2021} found that the assumption of a two parameter power-law continuum in addition to the emission line parameters (as used in the \citealt{Greig:2017a} model) breaks down when extrapolated to estimate the amplitude of the \lya{} forest continuum. However, for this study we are primarily interested in the \lya{} + \nv{} line region. Therefore, when comparing the various pipelines, we restrict the comparison to $>1208$~\AA.

\section{Example recovery of individual QSOs} \label{sec:qualitative}

After outlining the methodology of this blind QSO reconstruction challenge and the participating \lya{} reconstruction pipelines, we now explore the results. Firstly, we focus on providing visual examples, discussing the overarching qualitative performance of the various pipelines. Note, we cannot provide a visual summary of each QSO for each \lya{} reconstruction pipeline, thus we select a couple of examples. However, all results summarising this comparison can be found on a dedicated project page\footnote{\url{https://bradgreig.github.io/blind-QSO-challenge/}}.

For this visual comparison, we split the results firstly by sample (i.e. either X-Shooter or SDSS) in order to be able to separate out discussions related to the observational quality of the QSOs to be reconstructed. Secondly, for both blind samples we consider two types of objects, one with a prominent \lya{} emission line and with a relatively weak \lya{} line to distinguish performance based on the properties of the QSOs. Finally, for completeness, we provide a couple of examples of QSOs for which the majority of pipelines had issues providing a reasonable reconstruction.

\subsection{X-Shooter (Sample 1)}

\begin{figure*} 
	\begin{center}
	  \includegraphics[trim = 0.5cm 1.3cm 0cm 0.5cm, scale = 0.86]{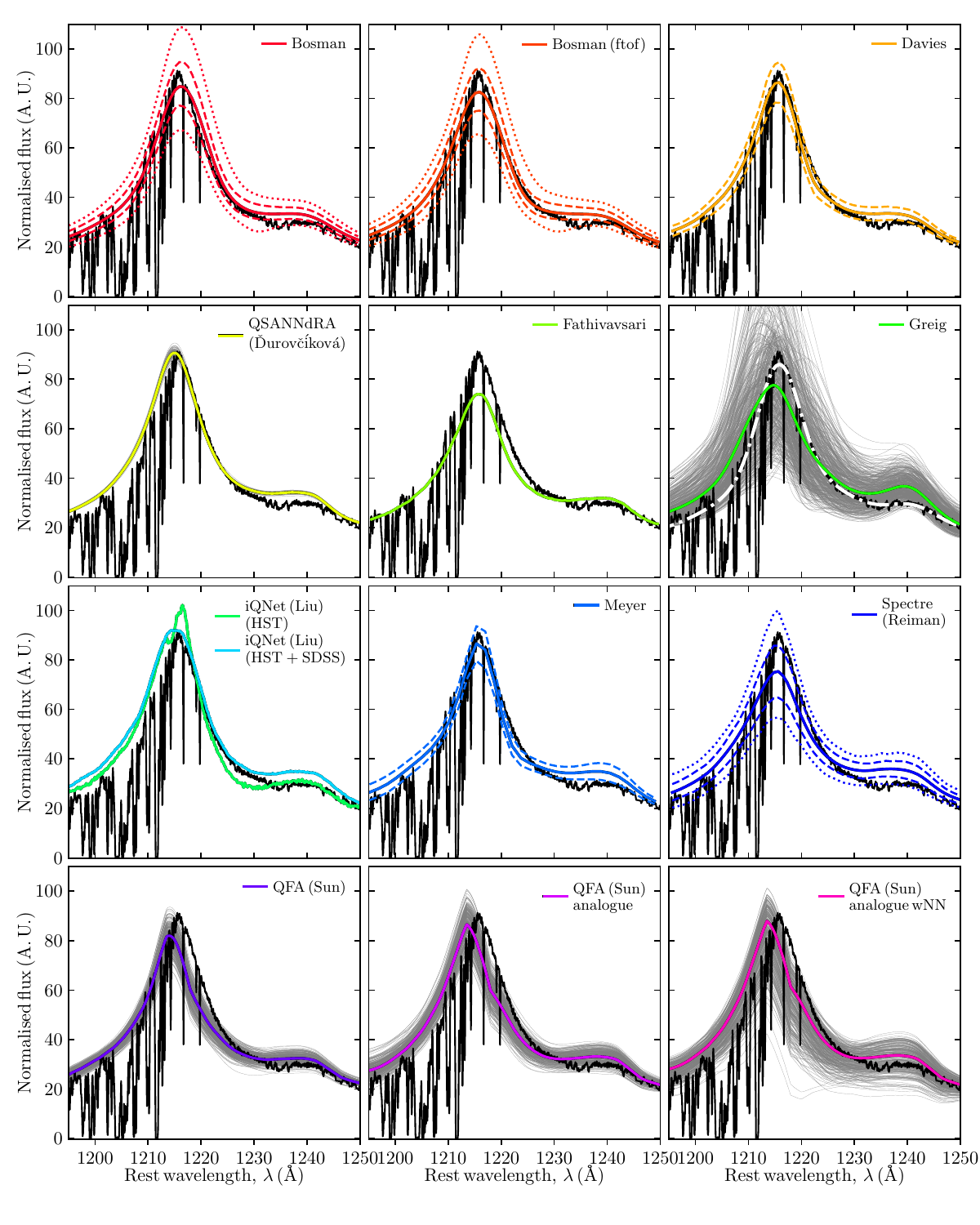}
	\end{center}
\caption[]{An example QSO (J0117+1552, $z=4.2428$) from the X-Shooter sample, representative of a QSO with a prominent \lya{} peak. Each panel corresponds to the predicted QSO from each QSO reconstruction pipeline. Some pipelines provide uncertainties as 68th and 95th percentile regions, which are denoted as dashed and dotted lines, respectively. For the Greig and QFA pipelines, thin grey lines correspond to random draws from the full posterior distribution. For the QSANNdRA pipeline, these correspond to 100 different (committee) neural networks (see text for further details). The white dot-dashed line in the Greig panel corresponds to the intrinsic QSO flux obtained by fitting the full QSO profile.}
\label{fig:Sample1a}
\end{figure*}

In Figure~\ref{fig:Sample1a} we provide the first reconstruction example with a prominent \lya{} emission line. Each panel corresponds to an individual reconstruction pipeline, with the coloured solid lines denoting the best-guess prediction. Dashed and dotted lines correspond to provided 68th and 95th percentile uncertainties on the QSO flux. For the Greig and QFA pipelines, the thin grey lines correspond to random draws from the full posterior distribution while for QSANNdRA the thin grey lines correspond to the 100 different trained neural network models (i.e. the committee, obtained using bootstrap aggregating whereby the same base machine learning architecture is re-trained 100 different times to improve the stability and accuracy). Finally, the white dot-dashed curve in the Greig panel corresponds to the intrinsic ground truth, obtained by MCMC fitting the full QSO spectrum.

For this QSO, almost all reconstruction pipelines perform equally well at predicting the full profile. Below we draw attention to several, object specific observations, and defer more detailed discussions and the implications of these to Section~\ref{sec:summary}. Interestingly, all pipelines over predict the intrinsic QSO flux at [1230, 1240]~\AA\ with varying degrees of severity. However, for those models that provide statistical uncertainties, this over prediction is typically near the 68th percent level, showcasing that within the statistical uncertainty the reconstruction pipelines agree with the intrinsic profile. In terms of the modelling uncertainty amplitude, the Greig model produces by far the broadest distribution. This stems from how the posterior distribution is sampled, that being from a 9 dimensional normal distribution that contains several emission line parameters that are not strongly correlated. These weaker line correlations result in a broader posterior distribution

With respect to the \lya{} line profile, several pipelines slightly under predict the true amplitude of the line, with the Fathivavsari, Greig and Spectre models being the most significantly affected. Additionally, for the QFA models, while not significantly under predicting the line amplitude, there is a slight offset (blueward) compared to the intrinsic profile (this is also true for the Greig pipeline, but to a lesser extent). However, we do not speculate on the origins of these observations at this point, as they only stem from a single object.

The dominant methodology in the literature for extracting QSO spectral features is via PCA decomposition, and one can see here that those based on PCA perform quite similarly (e.g. Bosman, Davies, QSANNdRA and Spectre). However, some differences do occur, either due to different training sets, differences in the pre-processing of the spectra and/or implementation (i.e. machine learning). For this particular object, the PCA methods perform better than the other approaches with respect to the best guess (i.e. maximum likelihood or machine learning prediction). Similarly the iQNet and Meyer pipelines equally perform well. While these machine learning approaches do not perform PCA decomposition they still decompose the QSO spectra into a low number of non-linear components through the usage of auto-encoders. Thus, all these approaches can be broadly classified as dimensionality reducing methods.

Amongst the QFA models, when switching from the default (full QSO population) model to one trained on a smaller subset of analogue QSOs the overall performance improves based on the \lya{} amplitude and shape (for this object, the blueshift remains). This is particularly interesting as all other reconstruction pipelines perform their training on large datasets in order to capture more variance in the QSO properties and assume that the learnt correlations are applicable to most types of QSOs. This is potentially something worth exploring in future for improving the overall performance of any of the existing reconstruction pipelines. However, care must be taken with such as approach as the smaller training sets may be more strongly affected by assumptions regarding the homogeneity of the analogue sample and the likelihood it accurately resembles the properties of the target QSO being reconstructed.

This X-Shooter object has a prominent \lya{} profile and is observed with higher resolution and S/N than the base BOSS training sets. Nevertheless, the various pipelines are able to recover this object relatively well, indicative that the BOSS training set contains QSOs representative of this type of object (or more specifically that it is a normal object). Further, both iQNet models (HST and HST+SDSS) perform quite similarly (albeit the HST model slightly overestimates the \lya{} amplitude) despite the fact that iQNet (HST) is trained on HST COS spectra, with notably higher spatial resolution than the X-Shooter object. Thus likely this object is also representative of the objects in the HST COS training set. 

\begin{figure*} 
	\begin{center}
	  \includegraphics[trim = 0.5cm 1.3cm 0cm 0.5cm, scale = 0.86]{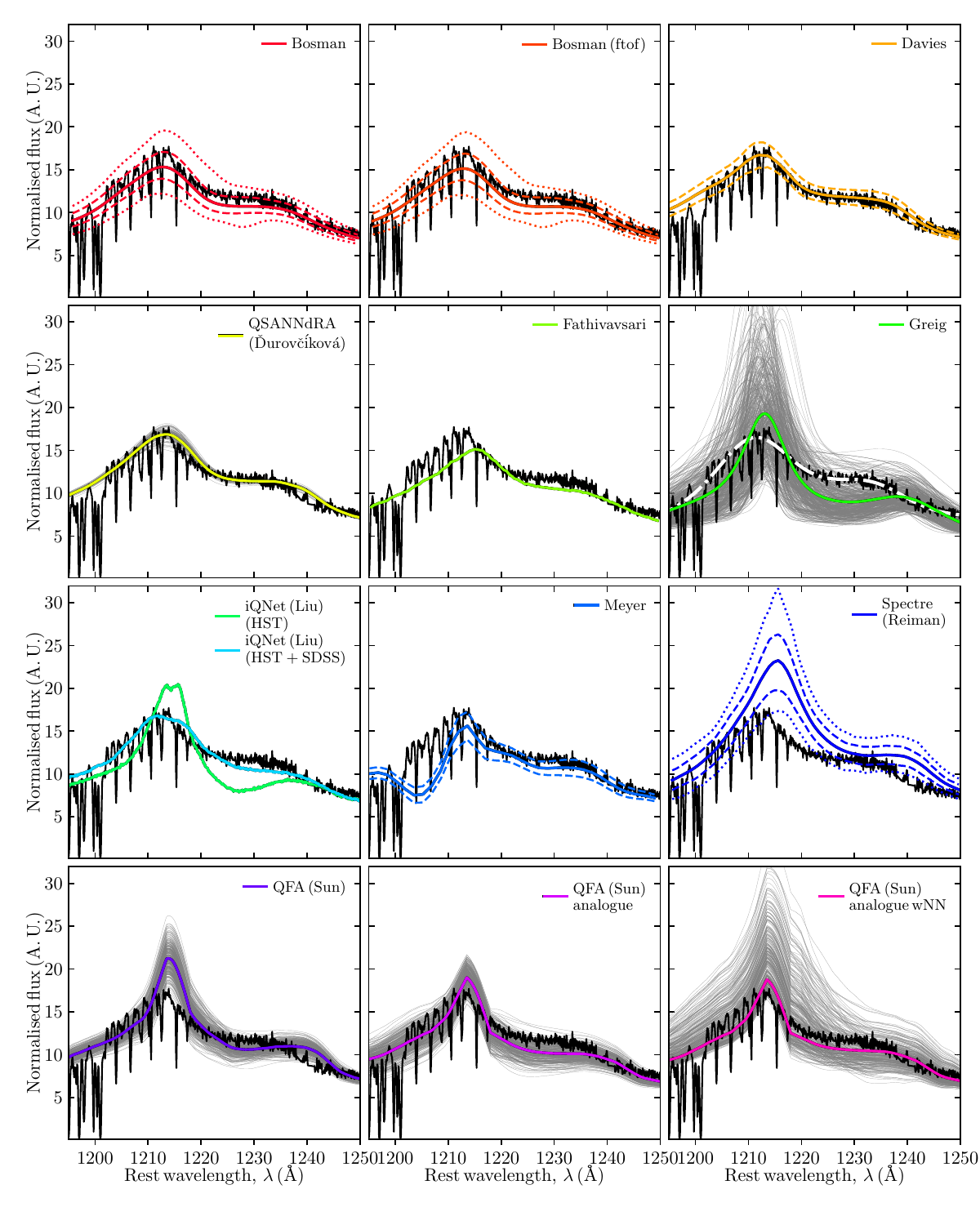}
	\end{center}
\caption[]{Same as Figure~\ref{fig:Sample1a} except for an X-Shooter QSO (J0100-2708, $z=3.5459$) with a much weaker \lya{} emission line profile.}
\label{fig:Sample1b}
\end{figure*}

In Figure~\ref{fig:Sample1b} we instead consider an X-Shooter spectra with much weaker \lya{} line emission. Here, a few larger discrepancies between the reconstructions and the intrinsic profile are apparent. Firstly, the Greig, iQNet (HST), Spectre and default QFA model all predict a sharper, more prominent \lya{} peak than demonstrated by the data. For iQNet (HST), this is simply due to the fact that this type of object with a weak \lya{} line is not represented by the training data, indicating that this pipeline may have issues reconstructing QSOs of this type. That is to say it is likely biased towards predicting profiles with sharper \lya{} profiles. However, this was not unexpected given our earlier discussions (see Section~\ref{sec:init-disc}). In particular, when BOSS data is included in the training set, the iQNet (HST+SDSS) model prediction is very close to the intrinsic profile, likely highlighting that it is indeed the characteristics of the spectra used in the training set driving the difference.

QFA learns the properties of median QSO population as a whole, which would expect slightly stronger peaked \lya{} lines. One can easily see this as for the analogue QFA models this overestimation of the peak goes away and the predictions are very close to the intrinsic truth. Again, this potentially highlights, where appropriate, the value of training on only representative objects rather than a full population as whole. For Spectre, these differences may simply be due to incorrect usage of the pipeline, as highlighted earlier.

Interestingly, the Greig pipeline predicts a very similarly shaped profile to that of iQNet (HST). For the Greig pipeline, these types of QSOs can be a little more problematic as the \lya{} line is predicted based off the expectation of a broad and narrow component. Further, the broad component can be strongly degenerate with the \nv{} line component, especially in the absence of a prominent \nv{} feature (as is the case here). The covariance matrix predictions are based off the line correlation strengths, which become much weaker when it is harder to disentangle the broad and narrow components. As a result, for these types of objects, this pipeline has a tendency to predict stronger \lya{} peaks, which is clearly evident here. This may indicate that the existing covariance matrix approach to describe the emission line parameters is not capturing all the available information and may require extending the model in future.

Like previously, the several different PCA or dimensionality reducing based methods perform very similarly given they are utilising very similar spectral information. In the case of the Meyer pipeline, there is a notable feature blueward of \lya{}, however, this may simply be due to an issue with the input data which produces a `feature' that is problematic for this approach. Given the goal was to perform a blind reconstruction, these authors had no knowledge of this at the time. Had the data been unblinded, this would have been identified and corrected. This is one potential downside of performing this blind comparison, that minor problems could result in spurious and incorrect predictions which would otherwise have been identified and fixed. Therefore, this needs to be kept in mind when analysing these results.

\subsection{SDSS sample (Sample 2)}

\begin{figure*} 
	\begin{center}
	  \includegraphics[trim = 0.5cm 1.3cm 0cm 0.5cm, scale = 0.86]{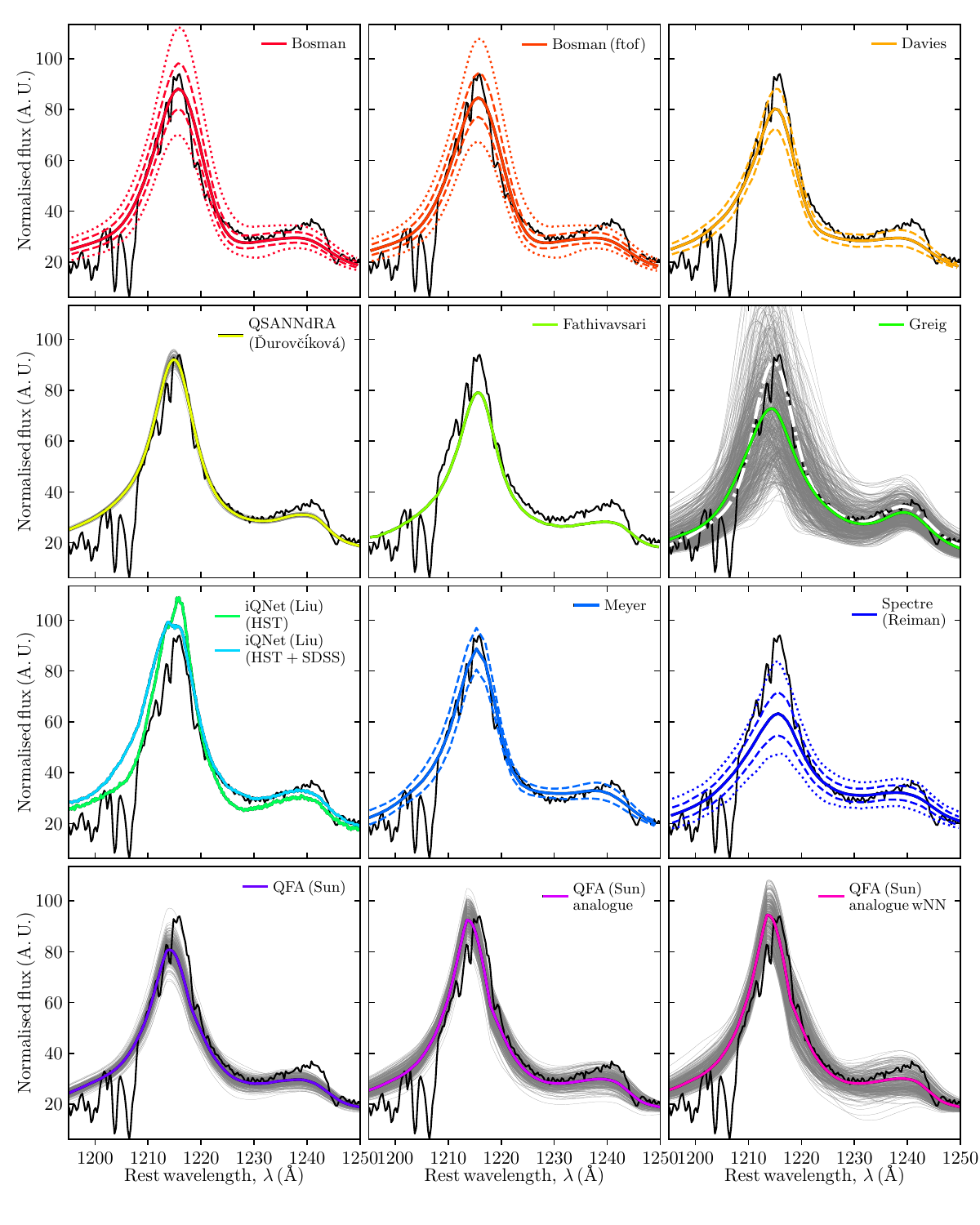}
	\end{center}
\caption[]{An example QSO (5184-56352-692, $z=3.64$) from the SDSS sample (Sample 2), representative of a QSO with a prominent \lya{} peak. Each panel corresponds to the predicted QSO from each QSO reconstruction pipeline. Some pipelines provide uncertainties as 68th and 95th percentile regions, which are denoted as dashed and dotted lines, respectively. For the Greig and QFA pipelines, thin grey lines correspond to random draws from the full posterior distribution. For the QSANNdRA pipeline, these correspond to 100 different neural networks (ensemble). The white dot-dashed line in the Greig panel corresponds to the intrinsic QSO flux obtained by fitting the full QSO profile.}
\label{fig:Sample2a}
\end{figure*}

\begin{figure*} 
	\begin{center}
	  \includegraphics[trim = 0.5cm 1.3cm 0cm 0.5cm, scale = 0.85]{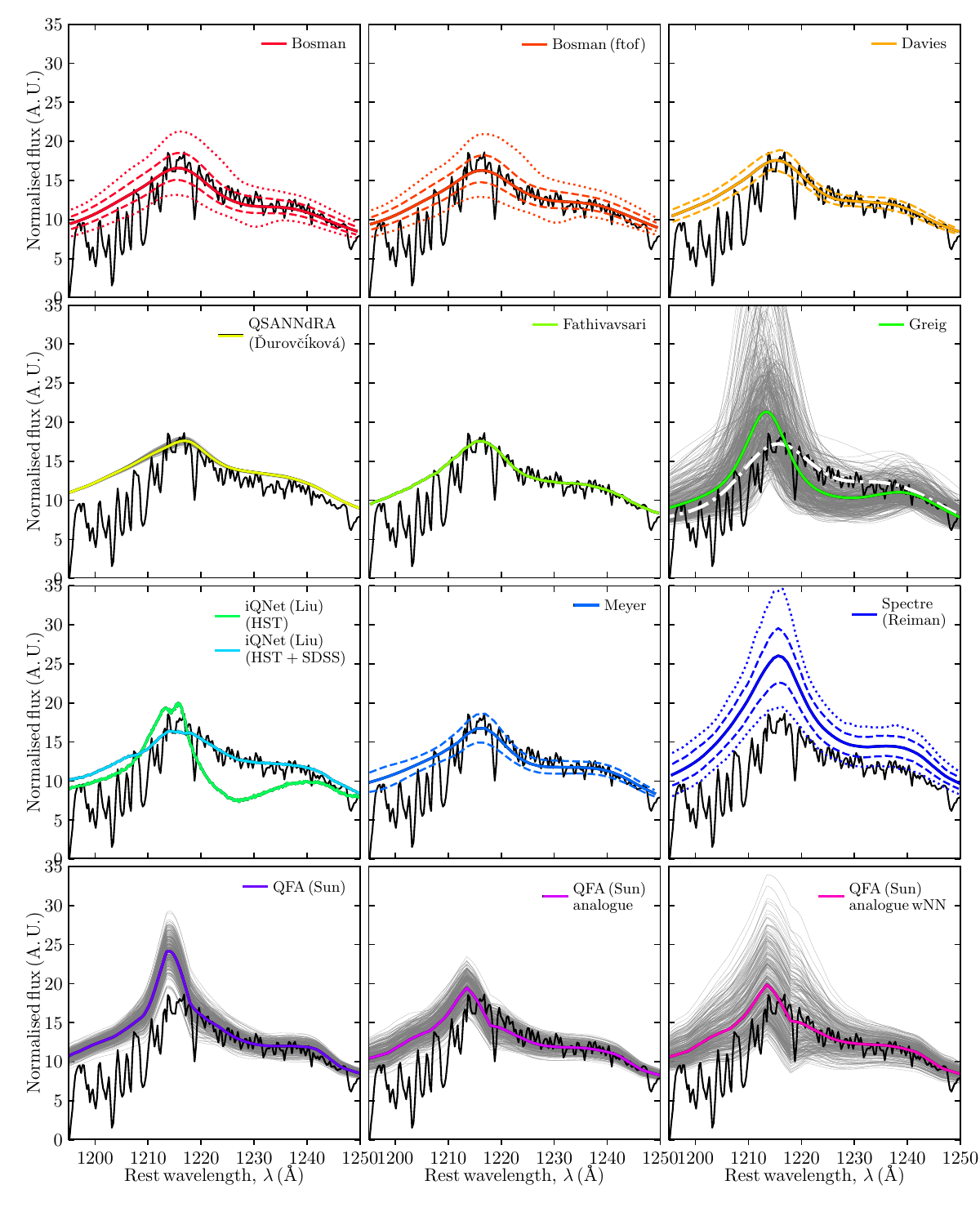}
	\end{center}
\caption[]{Same as Figure~\ref{fig:Sample2a} except demonstrating an SDSS QSO (5443-56010-898, $z = 3.69$) without a prominent \lya{} line profile.}
\label{fig:Sample2b}
\end{figure*}

We perform the same qualitative analysis as previously, except now on QSOs drawn from SDSS. In Figure~\ref{fig:Sample2a} we again provide an example of a QSO with a prominent \lya{} line. Interestingly, the same trends as seen for Figure~\ref{fig:Sample1a} are evident here. Firstly, again there are issues predicting the \nv{} line, this time with almost all pipelines under predicting the amplitude of the profile (whereas it was an over prediction previously). But also the shape of the \nv{} line is quite noticeably different. Unlike previously though, the \nv{} line is generally outside of the 68th percentiles, indicating this object is potentially more discrepant. This might allude to some issues predicting the \nv{} from redward information (i.e. $\lambda\gtrsim1260$\AA). In particular, \citet{Greig:2022} showed the correlation matrix of emission line properties and find \nv{} to correlate much more weakly than the other emission lines, which potentially could be leading to larger uncertainties with regard to predicting this line.

Again, the Fathivavsari, Greig and Spectre pipelines have similar issues predicting the amplitude of the \lya{} peak. Further, the QFA model again appears to be slightly blueshifted, but to a much lesser extent this time and improves on the overall amplitude determination. For this particular object, the PCA pipelines of Bosman and Davies this time slightly under predict the amplitude of the \lya{} peak, whereas QSANNdRA, iQNet and Meyer do not have such an issue. This potentially indicates that these machine learning based approaches might be more robust in reconstructing the diverse properties present in the QSO population owing to either the enlarged number of PCA components used (e.g. QSANNdRA) and/or the decomposition into a low number of non-linear components (e.g. iQNet and Meyer) which better characterise the broad diversity of QSO spectra.

In Figure~\ref{fig:Sample2b} we provide an example of an SDSS QSO with substantially weaker \lya{} profile. Again, the same observed trends for the X-Shooter example are seen for this SDSS QSO. The Greig, iQNet (HST), Spectre and original QFA model all predict a much more sharply peaked \lya{} emission line profile, which again is likely due to the features present in the training set. Trained on analogue QSOs rather than a large QSO distribution, the QFA analogue models perform well, consistent with all other reconstruction pipelines. This could also indicate that the full QFA model may benefit from the expansion of the number of components. Note, previously a spurious feature appeared for the Meyer pipeline, which is absent here, lending credence to the fact that it was likely driven by a quirk in the blind data rather than anything to do with the reconstruction methodology itself. 

\subsection{Example fails}

\begin{figure*} 
	\begin{center}
	  \includegraphics[trim = 0.5cm 1cm 0cm 0.5cm, scale = 0.50]{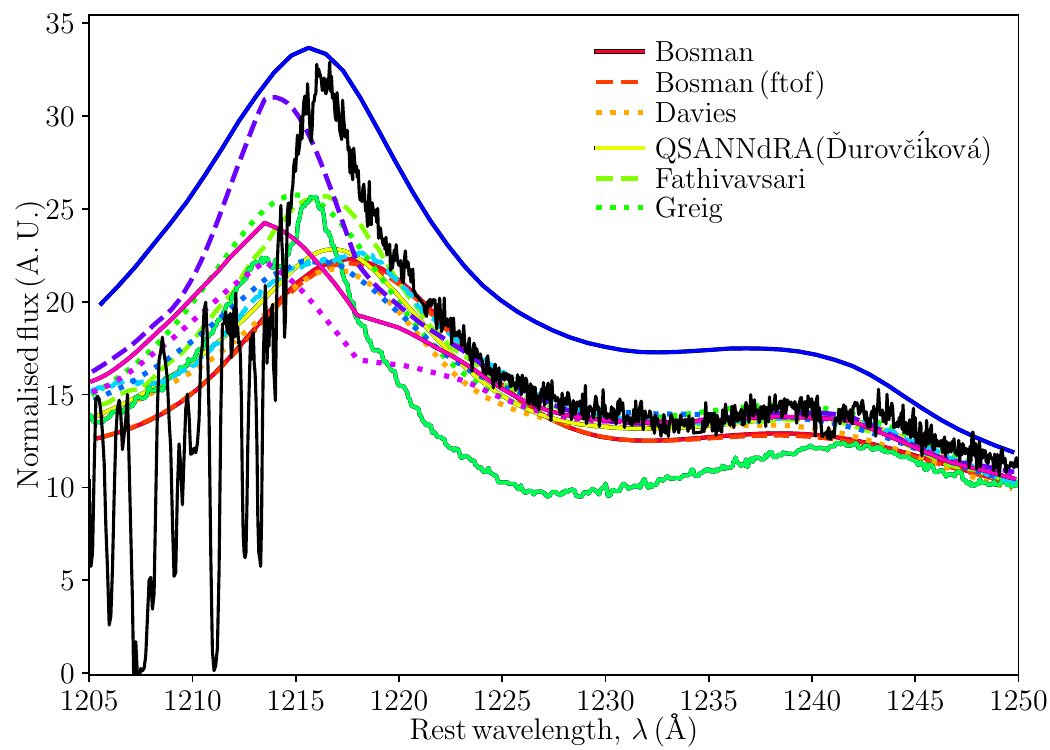}
	  \includegraphics[trim = 0.5cm 1cm 0cm 0.5cm, scale = 0.50]{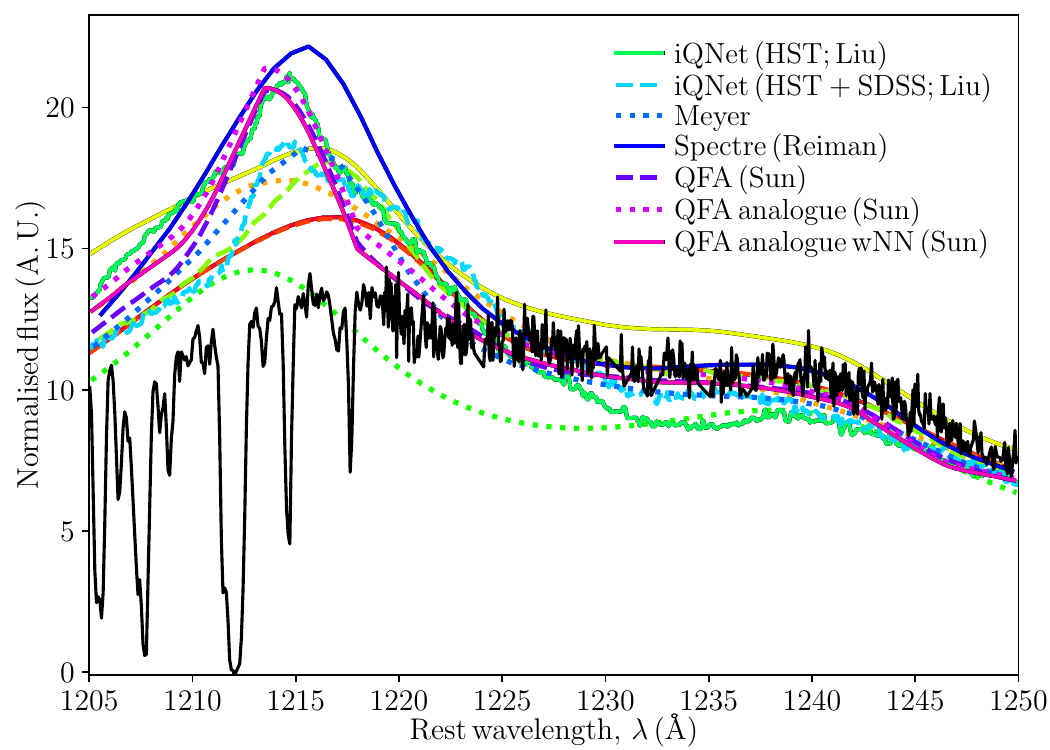}	
	\end{center}
\caption[]{Two examples of QSOs where the majority of reconstruction pipelines fail to accurately predict the \lya{} emission line profile, both from the X-Shooter sample (Sample 1). The left panel corresponds to J111701+131115 at $z=3.6218$, whereby the amplitude of the \lya{} emission is considerably under predicted. In the right panel, SDSSJ1416+1811 at $z=3.5928$, where the \lya{} emission is over predicted.}
\label{fig:Sample_fail}
\end{figure*}

Thus far we have focussed on providing examples where the reconstruction pipelines have performed well. Of course, this will not always be the case, and there are in fact several objects where the reconstructions are quite different from the intrinsic QSO profile. Each individual pipeline fails to reproduce a handful of objects over the entire sample, however these objects are not always the same for all pipelines. Obviously it is not possible to provide all examples of problematic reconstructions and equally it would be unfair to single out only a couple of pipelines as examples of failures. Note, in many of the cases of individual failure this could easily be caused by specific features in the provided blind QSOs data that produce erroneous predictions. Had the authors been able to inspect their \lya{} predictions relative to the intrinsic QSO spectrum these would easily be identified and rectified. As a result, there could potentially be a higher fraction of failures than would otherwise be the case. Therefore, here we aim to show a couple of representative examples of the typical types of failures, in cases where the vast majority of pipelines fail.

In Figure~\ref{fig:Sample_fail} we provide two examples of the most common types of failed predictions, both from the X-Shooter sample (Sample 1). However, these are not exclusive to the X-Shooter sample with a couple of these also occurring for Sample 2 (SDSS). In the left panel, which corresponds to the most common error, the \lya{} line profile amplitude and shape are considerably different for almost all pipelines. Note, the default QFA model performs relatively well for this object, however, the two QFA models trained on analogue QSOs to the redward spectral information present in this object do not, and perform equally poorly as all other pipelines. In this case, it is likely evident that this object is an outlier in the sense that its \lya{} emission is much stronger than its redward spectral information would predict. As an aside, here we have used the differences in the QFA methodology to assert the likelihood of this object being an outlier. One of the key values of QFA itself, explored in \citet{Sun:2022}, is that it can be used to identify outlier QSO based on the recovered properties relative to the QSO population used in the training set. Obviously this object could not have been identified as an outlier as it was blinded below 1260\AA. However, moving forward it will be useful to obtain pristine training sets of QSOs obeying certain properties.

In the right panel, we demonstrate an example of the opposite scenario where all reconstruction pipelines predict a much stronger \lya{} emission profile than is present in the observed spectra. The severity of this over prediction varies across the pipelines, but no one pipeline performs well. Interestingly, the three different QFA models predict comparable profiles indicating that the redward spectral information for this object is fairly typical of QSOs in general. Thus, this particular QSO exhibits a notably suppressed \lya{} profile than would otherwise be expected, likely indicating it is a bit of an oddball.

\subsection{Performance summary} \label{sec:summary}

\begin{figure*} 
	\begin{center}
	  \includegraphics[trim = 0.35cm 0cm 0cm 0.5cm, scale = 0.7]{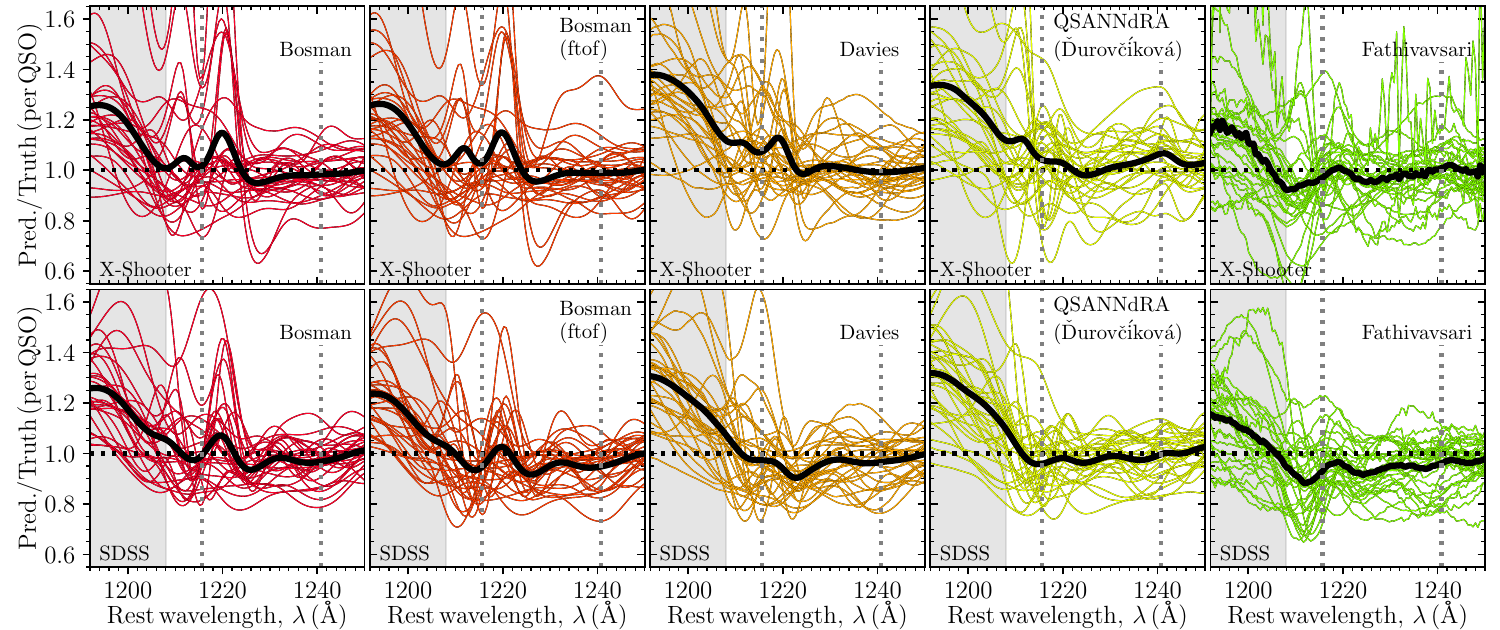}
	  \includegraphics[trim = 0.35cm 0cm 0cm 0cm, scale = 0.7]{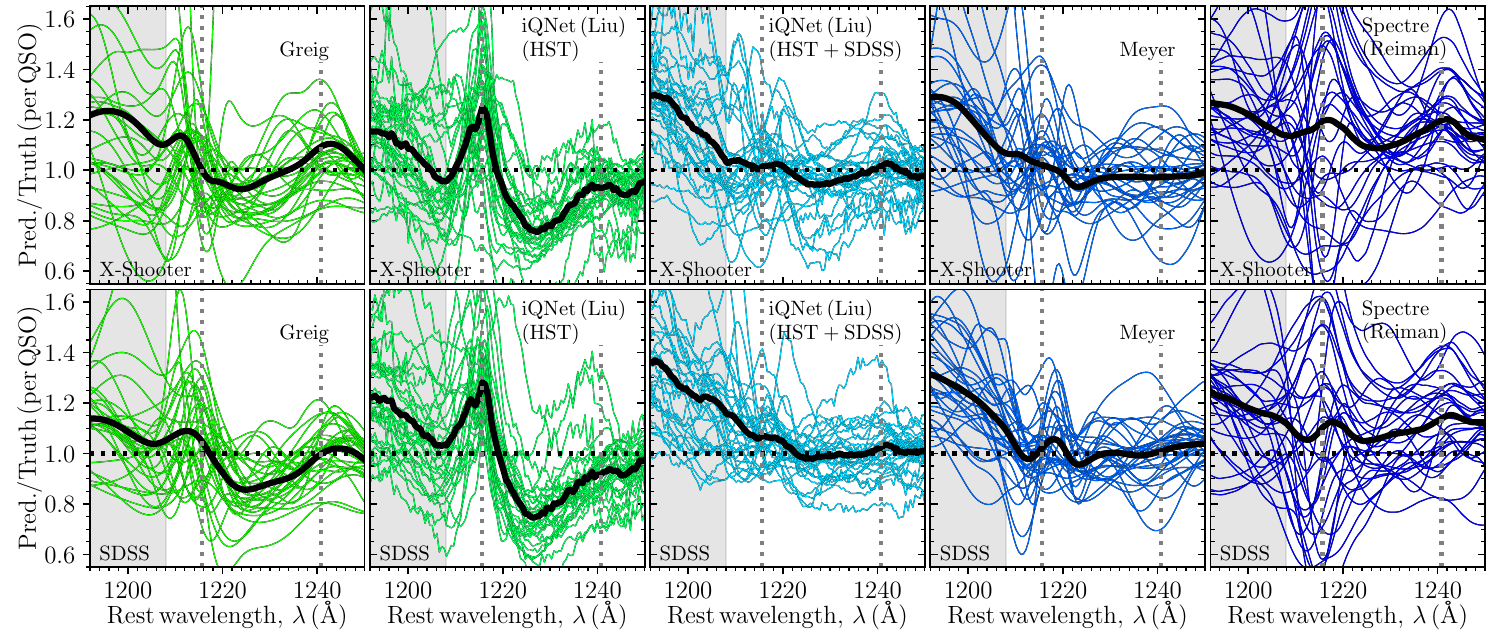}
	  \includegraphics[trim = 0.35cm 0.6cm 0cm 0cm, scale = 0.7]{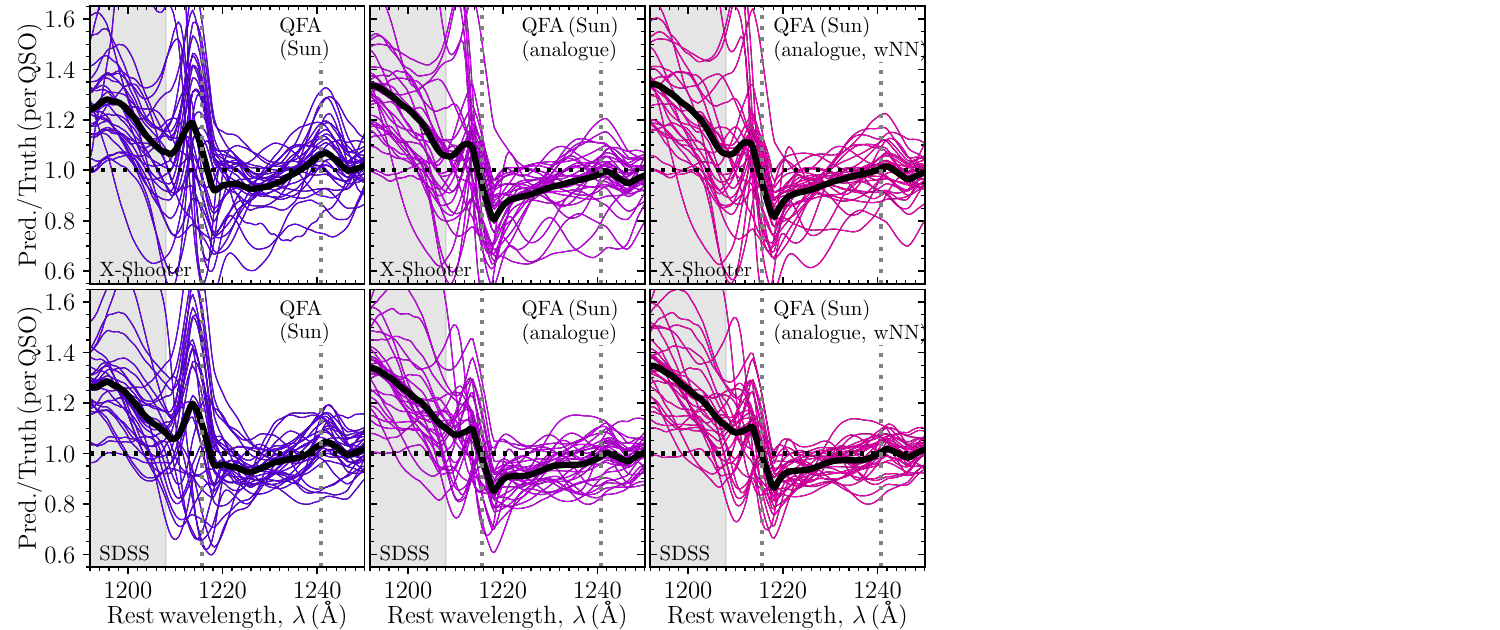}
	\end{center}
\caption[]{The ratio of predicted QSO flux to the true (intrinsic) QSO flux for each of the 30 QSOs in the blind QSO samples. Each QSO reconstruction pipeline appears in its own panel, distinguished by colour. The top (bottom) panel for each pipeline corresponds to the X-Shooter (SDSS) sample. The thick black curve in each panel corresponds to the mean flux ratio obtained over the entire sample of 30 QSOs in each sample for each pipeline. The vertical dotted grey lines correspond to rest-frame \lya{} and \nv{} respectively. The grey region at $\lambda < 1208$~\AA\ corresponds to where the intrinsic ground truth may not be as accurate owing to the determination method (see text for further details).}
\label{fig:Sample_means}
\end{figure*}

Since it is not possible to provide visual examples for all pipelines across the 60 different QSOs individually, we now provide a visual summary of the entire sample by presenting the ratio of predicted QSO flux by the intrinsic QSO flux. In Figure~\ref{fig:Sample_means}, we separate out the results for each pipeline individually (distinguished by colour) and additionally separate out by sample (top row X-Shooter and bottom row SDSS). The individual coloured lines correspond to the ratio of each individual QSO in the sample, whereas the thick black line corresponds to the mean ratio determined over the entire sample. For reference we denote the rest-frame locations of \lya{} and \nv{} by vertical grey dashed lines and the horizontal black dotted line corresponds to a ratio of unity (perfect reconstruction). Finally, the grey shaded region below $\lambda < 1208$~\AA\ corresponds to the cut-off where the intrinsic QSO flux may become less accurate owing to the power-law continuum used in the Greig model (see Section~\ref{sec:init-disc}). Note this choice is somewhat arbitrary, however, it is chosen to correspond to where the mean ratio (black curve) begins to diverge above unity at $\lambda < 1208$~\AA. Given we are primarily interested in the reconstruction of the \lya{} line and redward region out to \nv{} this does not impact our comparisons.

The key performance indicators to focus on in Figure~\ref{fig:Sample_means} are: (i) does the mean reconstruction (black curve) approach unity, in which case on average it is indicative of an unbiased reconstruction, (ii) is this mean relation featureless over the entire spectral region or does it depart from unity around a specific spectral location indicative of a systematic issue with the pipeline and (iii) the relative scatter in the individual ratios representative of the reconstruction performance over the sample with low scatter evident of a consistent performance across the sample or large scatter showing relatively poorer performance.

First and foremost, the mean performance of each QSO reconstruction pipeline is consistent across both samples (i.e. X-Shooter and SDSS). That is, each pipeline, trained on its particular training set, performs equally well (i.e. no systematic biases) irrespective of the quality and characteristics of the target QSO to be reconstructed (e.g. X-Shooter or SDSS). However, this only holds provided that the QSO training set is representative of the particular objects to be reconstructed. For example, iQNet (HST) exhibits large systematic offsets in the mean relation. This corresponds to consistently overestimating the \lya{} peak amplitude and underestimating the amplitude between \lya{} and \nv{}. This however was to be expected as the training set consists of HST COS spectra at $z<1$. The considerably higher resolution of these objects (along with the potential of these QSOs being a different population from the SDSS/X-Shooter QSOs) systematically predicts narrower and sharper emission line features than present in the X-Shooter and SDSS samples. Indeed, we saw this in Figures~\ref{fig:Sample1a}-\ref{fig:Sample2b}, where for the prominent \lya{} profiles there was a slight overestimate of the peak amplitude and slight underestimate of the flux between \lya{} and \nv{} (due to reduced blending of these lines in the higher resolution HST spectra). However, this was more striking in the weaker \lya{} line QSOs, which significantly overestimated the \lya{} profile. Interestingly, for iQNet, the mean relation for the X-Shooter and SDSS sample are consistent. Therefore, it appears to perform equally for both samples rather than favouring one over the other. Therefore, the relative differences between the X-Shooter and SDSS QSO samples are smaller than the differences between HST and either of the other two samples. This is further highlighted by the complete removal of these systematic offsets in the equivalent iQNet (HST+SDSS) sample. Therefore, when performing \lya{} reconstructions, one needs to be careful that the quality of the training set QSOs is comparable and more importantly contains representative objects to the target QSOs to be reconstructed. Too vast of a difference, could result in systematic biases.

The Greig pipeline also exhibits a systematic offset, but with a different shape. On average, the Greig pipeline underestimates the predicted QSO flux within the region between \lya{} and \nv{} by $\sim10-15$ per cent. Inspecting the individual profiles, it is clear that this trend is systematic with the vast majority demonstrating this decrement. The origin of this is the assumed Gaussian emission line modelling. The \lya{} line is modelled by a two component Gaussian (broad and narrow) along with a single component \nv{} line. The predicted QSO flux between the \lya{} and \nv{} line is then the sum of the broad \lya{} line and the \nv{}. If the \lya{} line is predicted to be bluer, the \nv{} line redder or either line to be narrower than it should be, then the sum of the flux within this region will be lower. The slight excess of the mean relation blueward of \lya{} and redward of \nv{} would tend to indicate that this is happening. Importantly, given that this approach performs well at fitting the intrinsic QSO flux, this may indicate that beyond first-order (Gaussian) line correlations may be required that are being missed by the current covariance matrix. Note this is the only pipeline that attempts to explicitly model the emission line profiles, indicating that spectral decomposition (e.g. PCA or factor analysis) is more robust at predicting the \lya{} line profile. Given this systematically occurs across all the reconstructed QSOs, in principle it can be corrected for. One could simply multiply the predicted \lya{} profiles by random draws of these templates and it would minimise this systematic bias from the full distribution at the cost of a marginally increased modelling uncertainty. This approach to correct for this systematic offset was adopted for the recent XQR-30 damping wing study (Greig et al., in prep).

For the default QFA model, we see an excess in the predicted flux near \lya{}. The individual profiles replicate this, with a large number demonstrating an excess, however, also several demonstrate a strong decrement. Earlier, we established the source of this, with the original QFA model appearing to overestimate the \lya{} flux in objects with weak to little \lya{} flux while underestimating the flux in QSOs with prominent \lya{} peaks. As highlighted earlier, the predictions improved once training sets of analogue objects were used. Within these QFA analogue models, one can see considerably less individual objects with such extreme predictions around \lya{}, although there are still several notable outliers, but these are relatively consistent with the number of outliers observed by the other pipelines. The transition from positive to negative for the mean profile through \lya{} can be explained by slight differences in the location of the \lya{} peak. As noted earlier, some of the visual examples demonstrated the predicted \lya{} peak to be slightly blue shifted. A tendency to produce bluer peaks would manifest as an excess blueward and a decrement redward. Thus the recovered shape around \lya{} demonstrated here likely points to a marginal preference for slightly blueward \lya{} peaks across the whole sample. Likely this can simply be corrected for in further iterations of this method.

For Spectre we see a systematic offset, along with very large scatter. Although the source is not immediately obvious, it likely signals a misunderstanding in the usage of Spectre for this comparison. Nevertheless, we keep all results for posterity, but caution against the over interpretation of its relatively poorer performance.

The remaining pipelines (e.g. Bosman, Davies, QSANNdRA, Fathivavsari, iQNet (HST+SDSS) and Meyer) all show fairly consistent performance, exhibiting essentially flat mean profiles over the entire spectral coverage, thus demonstrating more robust performance. Of these, the machine learning approaches QSANNdRA, Fathivavsari, iQNet (HST+SDSS) and Meyer tend to exhibit the strongest performance. The two Bosman pipelines exhibit a small bump in excess of unity near \lya{}, potentially indicative of a slight tendency to overestimate the \lya{} line. However, looking at the individual profiles, this is not the case and instead it appears to be most likely driven by having a slightly higher fraction of QSOs with an overestimate of the \lya{} profile driven by either a marginally broader predicted profile width or slight redshift of the line centre. To a lesser extent, the Davies pipeline also exhibits this behaviour but only for the X-Shooter sample, and this may also be true for the Fathivavsari pipeline. What is most interesting about these better performing pipelines is that they are all machine learning based approaches either directly using PCA decomposition (QSANNdRA) or dimensionality reduction of the QSO spectra (Meyer, iQNet and Fathivavsari). Therefore potentially providing evidence for a preference for PCA or dimensionality reducing approached for better \lya{} profile reconstruction performance. Additionally, the updated QFA models also show considerable promise at performing comparable to these approaches, however this is not too surprising given the analogies between PCA and factor analysis. Thus, in general the better performing reconstruction pipelines are those utilising some form of spectral decomposition to more robustly extract the relevant information from the QSO spectrum.

Generally speaking, our primary interest in these \lya{} reconstruction pipelines is for exploring the IGM damping wing within the spectra of $z\gtrsim6$ QSOs. Within the literature, these studies can be broken down into two broad approaches. One focussing only redward of \lya{} \citep[$\gtrsim1220$\AA, e.g.][]{Greig:2017,Greig:2022} whereas others extend to various degrees blueward of \lya{} \citep[e.g.][]{Mortlock:2011p1049,Banados:2018,Davies:2018,Dominika:2020,Reiman:2020,Wang:2020,Yang:2020}. As shown in Figure~\ref{fig:Sample_means}, the relative differences within these regimes can be quite different. It is considerably more difficult to accurately model the \lya{} line, as evident by the larger scatter and presence of systematic offsets in the mean \lya{} predictions. On the other hand, redward of \lya{} there is considerably less scatter indicating much more robust performance. However, ultimately there is a tradeoff here as the IGM damping wing signal becomes progressively weaker redward of \lya{}, therefore although the relative scatter and flux differences are smaller, they still may be larger than the signal and associated uncertainties with modelling and extracting the damping wing. Therefore, a decision based on relative pipeline performance might be dependent on the specific application. Although in general the better performing pipelines perform equally well within both regimes, for example the QFA models have relatively small scatter redward of \lya{}, indicating these might perform well for redward damping wing studies.

\begin{figure*} 
	\begin{center}
	  \includegraphics[trim = 0.1cm 1.2cm 0cm 0.4cm, scale = 0.94]{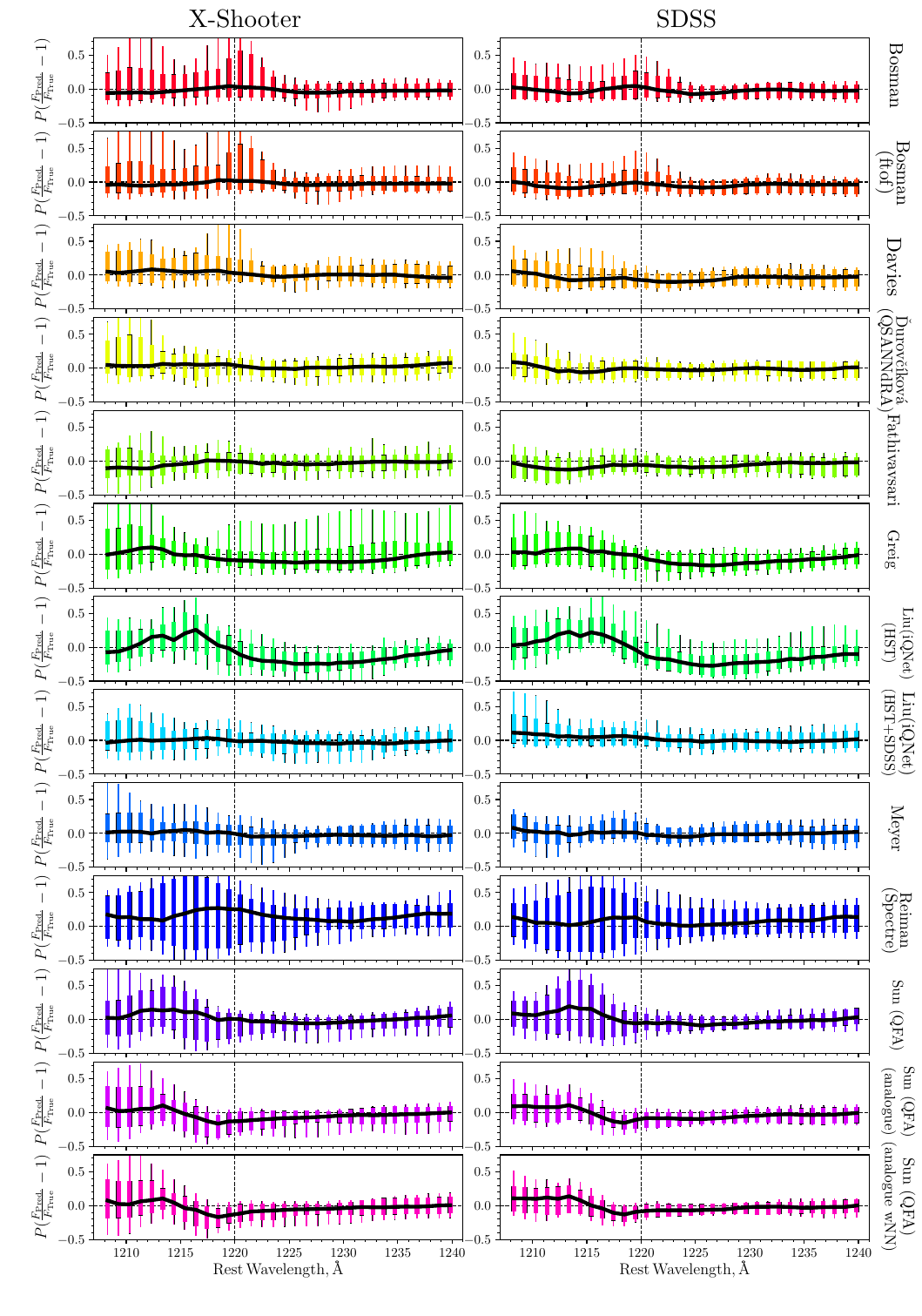}
	\end{center}
\caption[]{The PDF of predicted QSO flux relative to the true QSO flux determined over the entire QSO sample (X-Shooter; right and SDSS; left) as a function of wavelength. Specifically, we only consider the range [1208, 1240]~\AA, with the lower bound corresponding to where intrinsic continuum determination becomes less accurate. We provide this PDF for all QSO reconstruction pipelines, with the thick bars and thin lines denoting the 68th and 95th percentiles respectively while the black line with white diamonds represent the median.}
\label{fig:perbin-rel}
\end{figure*}

\begin{figure*} 
	\begin{center}
	  \includegraphics[trim = 0.1cm 1.2cm 0cm 0.65cm, scale = 0.69]{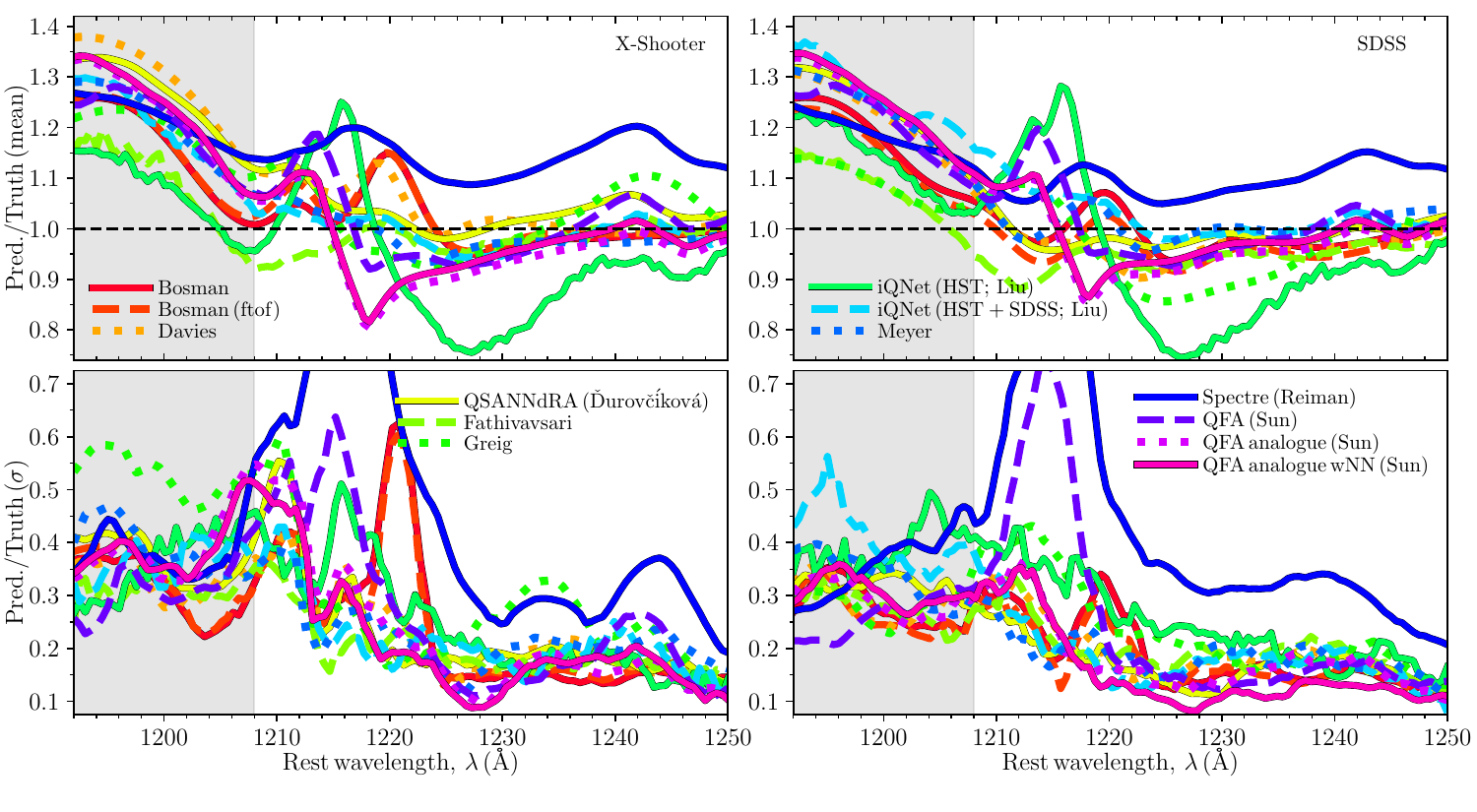}
	\end{center}
\caption[]{A summary of the flux ratio (prediction over truth) for each of the different QSO pipelines. The top row corresponds to the mean flux ratio while the bottom row corresponds to the 68th percentiles on this flux ration determined from the 30 QSOs in each sample. The left panel corresponds to the X-Shooter sample and the right corresponds to SDSS. The horizontal black dashed line at unity indicates the goal of achieving perfect unbiased reconstruction performance.}
\label{fig:MeanSigma}
\end{figure*}

To more robustly compress the information from these observations and the individual profiles shown in Figure~\ref{fig:Sample_means}, in Figure~\ref{fig:perbin-rel} we compute the probability distribution function (PDF) of the predicted QSO flux compared to the estimated intrinsic QSO flux for each QSO sample and reconstruction profile as a function of wavelength. Here, we specifically only consider the spectral range from [1208, 1240]~\AA, where the lower bound comes from where we deem our intrinsic determination method to begin to be less accurate. The thick bars correspond to the 68th percentiles and thin lines are the 95th percentiles while the black curve with white diamonds corresponds to the median. This enables a clearer visual comparison of the relative scatter and offsets within each reconstruction panel. To accompany this, we also provide Figure~\ref{fig:MeanSigma}, which over plots the mean prediction divided by intrinsic profile determined over each sample of 30 blind QSOs (black curves from Figure~\ref{fig:Sample_means}) along with the associated 68th percentile scatter in the 30 individual predictions for each sample. These figures provide a clearer demonstration of the relative amplitudes of the mean pipeline performance and overall scatter. One can clearly see that at $\lesssim1220$\AA, there are large systematic offsets in various reconstruction pipelines, along with much larger amplitude scatter indicating that one must be much more careful in the choice of \lya{} reconstruction pipeline when working in this regime. On the other hand, at $\gtrsim1220$\AA, the vast majority of pipelines converge in regards to both mean performance ($<10$ per cent offset) and the scatter reduces by a factor of $2-3$. 

\section{Statistical characterisation of performance} \label{sec:quantitative}

Previously, we explored individual \lya{} predictions for the blind QSOs, extracting several qualitative insights about the performance of the various reconstruction pipelines. In this section, we endeavour to perform a more quantitative analysis. Firstly, we explore the distribution of best-guess predictions (e.g. maximum likelihood or maximum $a$-posteriori) to be inclusive of all available pipelines. Then, we focus only on those pipelines that statistically characterise their modelling uncertainties. Throughout, we will split this quantitative exploration both by sample (X-Shooter or SDSS) and by spectral region, [1208, 1220]~\AA\ and [1220, 1240]~\AA. This latter choice is to separate out performance over the \lya{} line region and redward of \lya{}, respectively.

\subsection{Best-guess predictions}

To quantify the performance of the various \lya{} reconstruction pipelines, our first metric of choice is to determine the distribution of QSOs whose predicted flux lies within a certain fractional threshold of the intrinsic flux over a specific wavelength region. Although this is somewhat arbitrary, this choice is adopted to both compress the available information over a wavelength region into a single value while down weighting reconstructions which either produce spurious, non-smooth features across neighbouring bins or are systematically biased within some narrow spectral region. Thus, the goal is to determine which pipelines produce realistic profiles and perform well at predicting the intrinsic flux over the entire wavelength region. If one was to attempt to average the information across the entire spectral range, it may not sufficiently down weight pipelines that produce systematic oscillatory features (e.g. the Greig pipeline which over and under estimates the intrinsic QSO flux within different spectral regions).

\begin{figure*} 
	\begin{center}
	  \includegraphics[trim = 0.3cm 0.7cm 0cm 0.5cm, scale = 0.635]{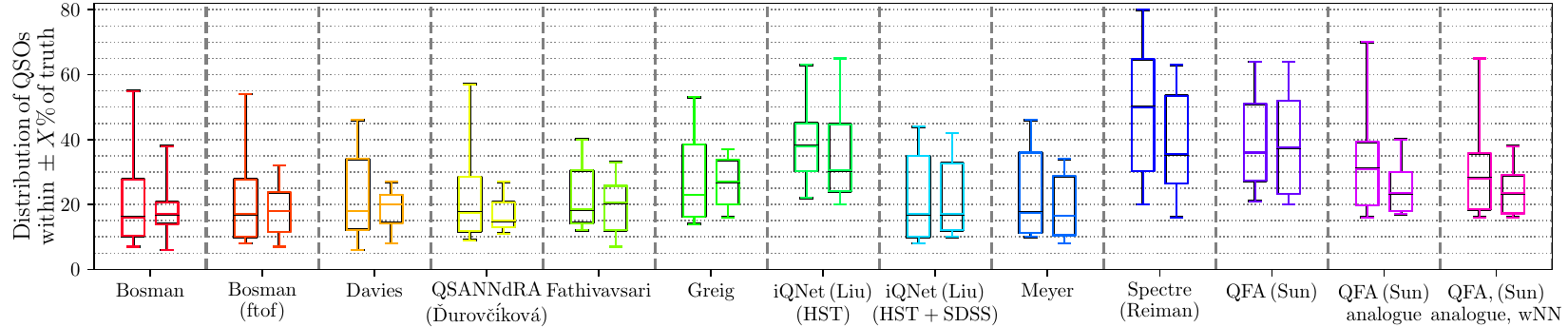}
	\end{center}
\caption[]{The distribution of QSOs for each sample whose predicted flux lies within a given percentage of the intrinsic QSO flux over the [1208, 1200]~\AA\ region. The box plot corresponds to the 25, 50, 75th percentiles (7, 15 and 22 QSOs within the given flux percentage) whereas the whiskers denote the 10 and 90th percentiles (3 and 27 QSOs). Each pipeline is separated by sample, with the left (right) box plot corresponding to the X-Shooter (SDSS) sample, respectively. The best performing pipelines appear along the bottom, corresponding to the most number of QSOs within the smallest possible flux percentage away from the intrinsic truth.}
\label{fig:Sum0820}
\end{figure*}

In Figure~\ref{fig:Sum0820}, we highlight the distribution of QSOs within each sample whose predicted flux is within a specific threshold of the intrinsic flux over the [1208, 1220]~\AA\ region. We represent this information as a box and whisker plot, with the box demonstrating the flux threshold containing the 25, 50, 75th percentiles (i.e. 7, 15 and 22 QSOs) and the whiskers denote the 10 and 90th percentiles (3 and 27 QSOs). Optimal performance is then demonstrated by the distribution centred around the lowest possible flux percentage and preferentially with the narrowest box width and whiskers (i.e. reduced scatter). Each individual pipeline is colour coded as previously and the X-Shooter sample appears on the left, with the SDSS sample on the right for each individual pipeline respectively. 

Firstly, almost all pipelines perform better for SDSS than for the X-Shooter sample, both in terms of the threshold that contains 75 per cent of the sample along with the reduction in box width (scatter). This is somewhat counter to our previous exploration where, based off individual profiles, there did not appear to be any obvious preference. Focussing on the 90th percentiles (upper whisker) it is clear these extend more for the X-Shooter than for the SDSS sample, potentially indicating that there are more outliers or spurious objects in the former sample. That is, this sample potentially includes more objects that are inconsistent with the characteristics of the SDSS training set, for instance the XQ-100 sample are known to have higher luminosities and black-hole masses. In Figure~\ref{fig:Sample_means}, it does appear that there are a higher fraction of QSOs with larger amplitude variations in the X-Shooter sample than the SDSS sample. Nevertheless, the overall difference in the threshold to contain the same fraction of QSOs is fairly modest, generally speaking a difference in flux threshold of $\lesssim10$ per cent between the two samples. Likely increasing the relative size of the blind QSO set would also diminish the differences. iQNet (HST) does not obey this trend, with a narrower scatter for the X-Shooter sample. However, the upper-envelope (75th percentile) remains the same for both, thus the reduced scatter comes due to the X-Shooter QSOs being slightly more discrepant from the intrinsic profile on average.

Under this metric, the best performing pipelines are both Bosman, QSANNdRA and Fathivavsari, with each containing 75 per cent of their QSO predictions within a 30 per cent flux threshold for both samples (20-25 per cent for SDSS only). Only marginally behind are Davies, iQNet (HST+SDSS), Meyer and QFA (analogue wNN) within a threshold of 35 per cent. Note however that both the iQNet (HST+SDSS) and Meyer pipelines have notably higher scatter, potentially indicating a larger fraction of objects that are not as robustly reconstructed. The Greig and QFA (analogue) pipeline are contained within a 40 per cent threshold. Once again, the dominant trend here is that the vast majority of the better performing pipelines are those that rely on spectrally decomposed information (PCA, factor analysis or auto-encoders).

\begin{figure*} 
	\begin{center}
	  \includegraphics[trim = 0.3cm 0.7cm 0cm 0.5cm, scale = 0.635]{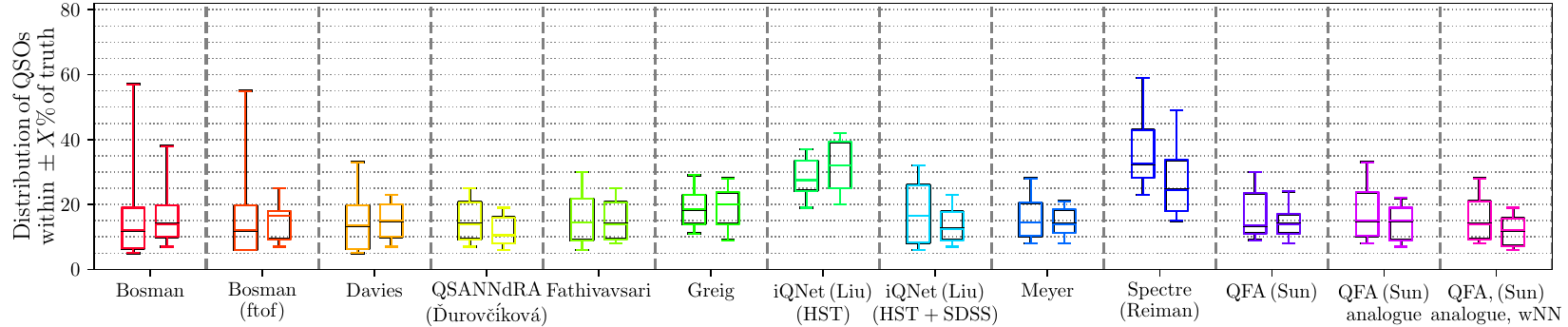}
	\end{center}
\caption[]{The same as Figure~\ref{fig:Sum0820} except over the [1220, 1240]~\AA\ wavelength range.}
\label{fig:Sum2040}
\end{figure*}

In Figure~\ref{fig:Sum2040} we focus instead on the information redward of \lya{}, [1220, 1240]~\AA. Unsurprisingly, the overall performance within this region improves with all but two pipelines containing 75 per cent of the QSOs within only a 20 per cent flux threshold. The two discrepant pipelines are iQNet (HST) and Spectre. For the former, the discrepancy is due to the HST training set, which prefers on average a more sharply peaked \lya{} profile which drops away faster between \lya{} and \nv{} (see Figure~\ref{fig:Sample_means}). For Spectre, we suspect unresolved issues resulting from our implementation of the pipeline.

Within this spectral region, the best performing pipelines are QSANNdRA and QFA (analogue wNN) which contain 75 per cent of the QSOs within 20 (15) per cent for the X-Shooter (SDSS) samples respectively. The Bosman, Davies, Fathivavsari and Meyer pipelines are only marginally behind with a slightly higher threshold for the SDSS sample. Increasing this flux threshold out to 25 per cent, picks up the Greig, QFA and QFA (analogue) pipelines. Note however, the Bosman pipelines contain larger whiskers (higher 90th percentiles) indicative of a higher number (larger than 3 QSOs) that are discrepant by over 50 per cent, which is driven by the broader distribution with larger positive tail (over estimation of the QSO flux) within around 1220\AA\ (see e.g. Figure~\ref{fig:perbin-rel}). Likely this is either a higher fraction of QSOs within the X-Shooter sample inconsistent with the respective training sets, or a particular feature in the blind QSO spectra that has lead to spurious results.

\subsection{Probability distribution}

For the most part, our comparisons have only focussed on the best-guess prediction of the \lya{} profile. However, what is more relevant are the associated uncertainties, machine learning versus full posteriors. Therefore, in this section we explore the performance of the reconstruction pipelines, folding in the provided statistical uncertainties. To do so, we retain the same metric as above, the determination of the distribution of QSOs whose predicted flux lies within some flux threshold of the intrinsic flux over a spectral region, but, instead of the flux threshold we adopt as a threshold the statistical uncertainty of the model prediction. Instead here, our threshold flux is a fraction of the statistical uncertainty. Remember, the logic for this metric was to penalise individual reconstructions which might perform well over most of the region but contain a discrepant (unphysical) feature otherwise. Note, both iQNet and the Fathivavsari models do not provide modelling uncertainties and thus are not considered.

\begin{figure*} 
	\begin{center}
	  \includegraphics[trim = 0.5cm 0.5cm 0cm 0.5cm, scale = 0.635]{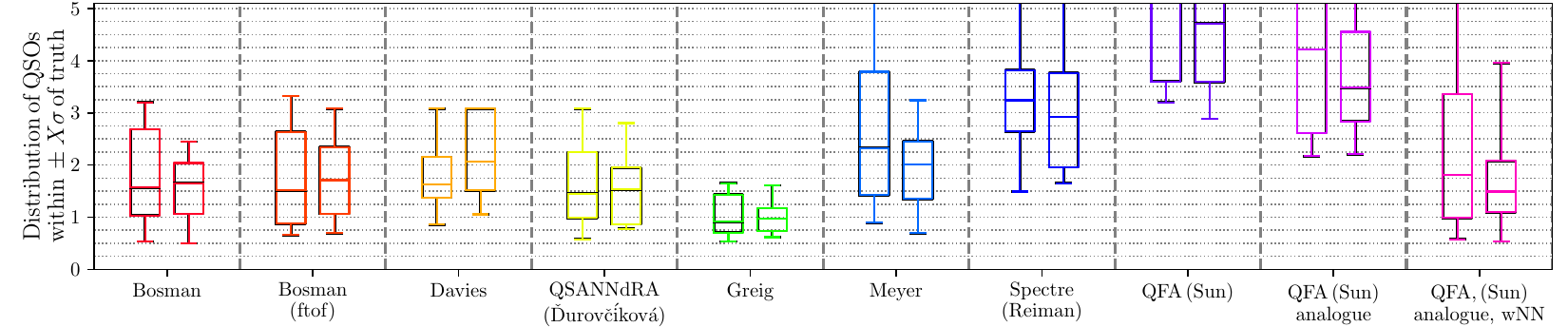}
	\end{center}
\caption[]{The distribution of QSOs in each sample whose predicted QSO flux is some $\pm X \sigma$ within the true intrinsic QSO flux over the [1208, 1220]~\AA\ region. The box plot corresponds to the 25, 50, 75th percentiles (7, 15 and 22 QSOs within the given flux value) whereas the whiskers denote the 10 and 90th percentiles (3 and 27 QSOs). Each pipeline is separated by sample, with the left (right) box plot corresponding to the X-Shooter (SDSS) sample, respectively. The best performing pipelines appear along the bottom, corresponding to the most number of QSOs within the smallest possible flux range away from the intrinsic truth. Note, this only includes pipelines capable of providing prediction uncertainties.}
\label{fig:Sum0820-err}
\end{figure*}

In Figure~\ref{fig:Sum0820-err}, we highlight the distribution of QSOs contained within some fraction of the modelling uncertainties across [1208, 1220]~\AA. Under this particular definition, the Greig pipeline performs best containing 90 per cent of the QSOs within $2\sigma$. However, this is unsurprising given that this pipeline has the largest modelling uncertainties owing to its methodology. Thus, it should always be easier to predict the QSO flux within a lower amplitude away from the intrinsic value as larger uncertainties will compensate for any systematic over/underestimate of the \lya{} line amplitude.

Interestingly, the second best performing pipeline is QSANNdRA, which has by far the smallest uncertainties. Unlike all others, QSANNdRA's uncertainties are not drawn from a statistical distribution characterising the scatter in the QSO training set population. Instead, they are simply a measure of the variance due to varying the initial conditions when training the committee (ensemble) of 100 artificial neural networks. Therefore in comparison to the other pipelines, QSANNdRA's provided uncertainty is not equivalent, and underestimates the true uncertainty within the QSO population. The fact that QSANNdRA still performs well under this metric highlights just how well this pipeline generally performs. 

Only marginally further behind QSANNdRA in terms of performance are the Bosman, Davies and QFA (analogue wNN) pipelines. The Meyer pipeline, is equivalent to these for the SDSS sample, but slightly worse for the X-Shooter sample where it appears to have a larger scatter in the reconstructed profiles (see Figure~\ref{fig:Sample_means}). The relatively poorer performance of QFA and the analogue QFA owe to a mix of relatively small modelling uncertainties (see Figures~\ref{fig:Sample1a}-\ref{fig:Sample2b}) and apparent slight blueshift in the location of the \lya{} profile. For the QFA (analogue wNN) this is a little less extreme as the modelling uncertainties have increased.

\begin{figure*} 
	\begin{center}
	  \includegraphics[trim = 0.5cm 0.5cm 0cm 0.5cm, scale = 0.635]{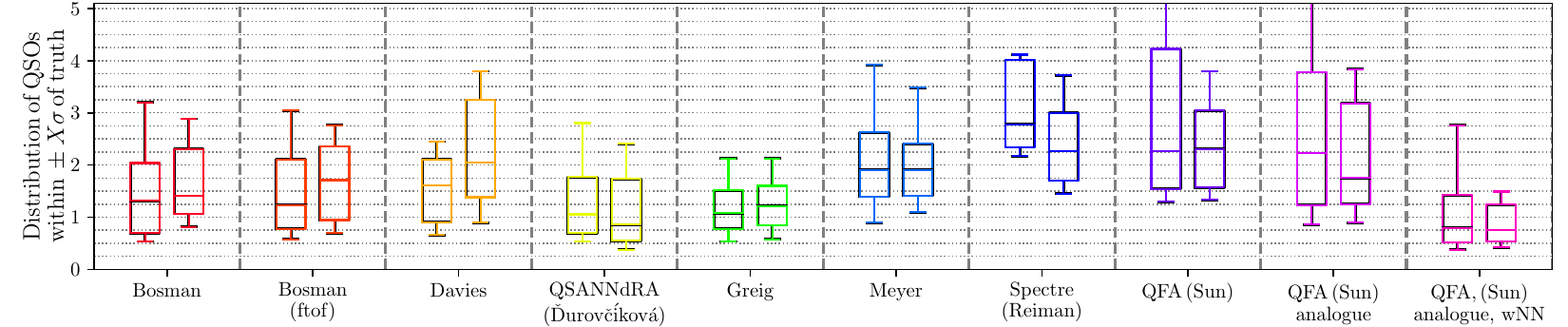}
	\end{center}
\caption[]{The same as Figure~\ref{fig:Sum0820-err} except over the [1220, 1240]~\AA\ wavelength range.}
\label{fig:Sum2040-err}
\end{figure*}

In Figure~\ref{fig:Sum2040-err}, we present the results over [1220, 1240]~\AA. Here, the three best performing pipelines are again Greig, QSANNdRA and QFA (analogue wNN), all with 75 per cent of the QSOs across both samples within $\sim2\sigma$. Both Bosman pipelines, Davies and Meyer are only marginally further behind. Within this spectral region the Greig pipeline is most severely affected by its systematic offset (see Figure~\ref{fig:Sample_means}), however the larger modelling uncertainties compensate for this resulting in comparable performance to the other pipelines. Once again, the relative strong performance of QSANNdRA, despite its much smaller uncertainties, highlights its overall strength. 

This metric of determining the distribution of QSOs within some fractional value of the modelling uncertainty tended to favour reconstruction pipelines with the largest errors. Here, we aim to more robustly quantify the reconstruction performance by taking into account both the \lya{} predictions and the relative amplitude of the uncertainties. We explore this in two ways: (i) determined over the entire sample as a function of wavelength and (ii) considering each QSO individually determined over the entire wavelength coverage of our reconstruction.

\begin{figure*} 
	\begin{center}
	  \includegraphics[trim = 0.8cm 0.5cm 0cm 0.35cm, scale = 0.96]{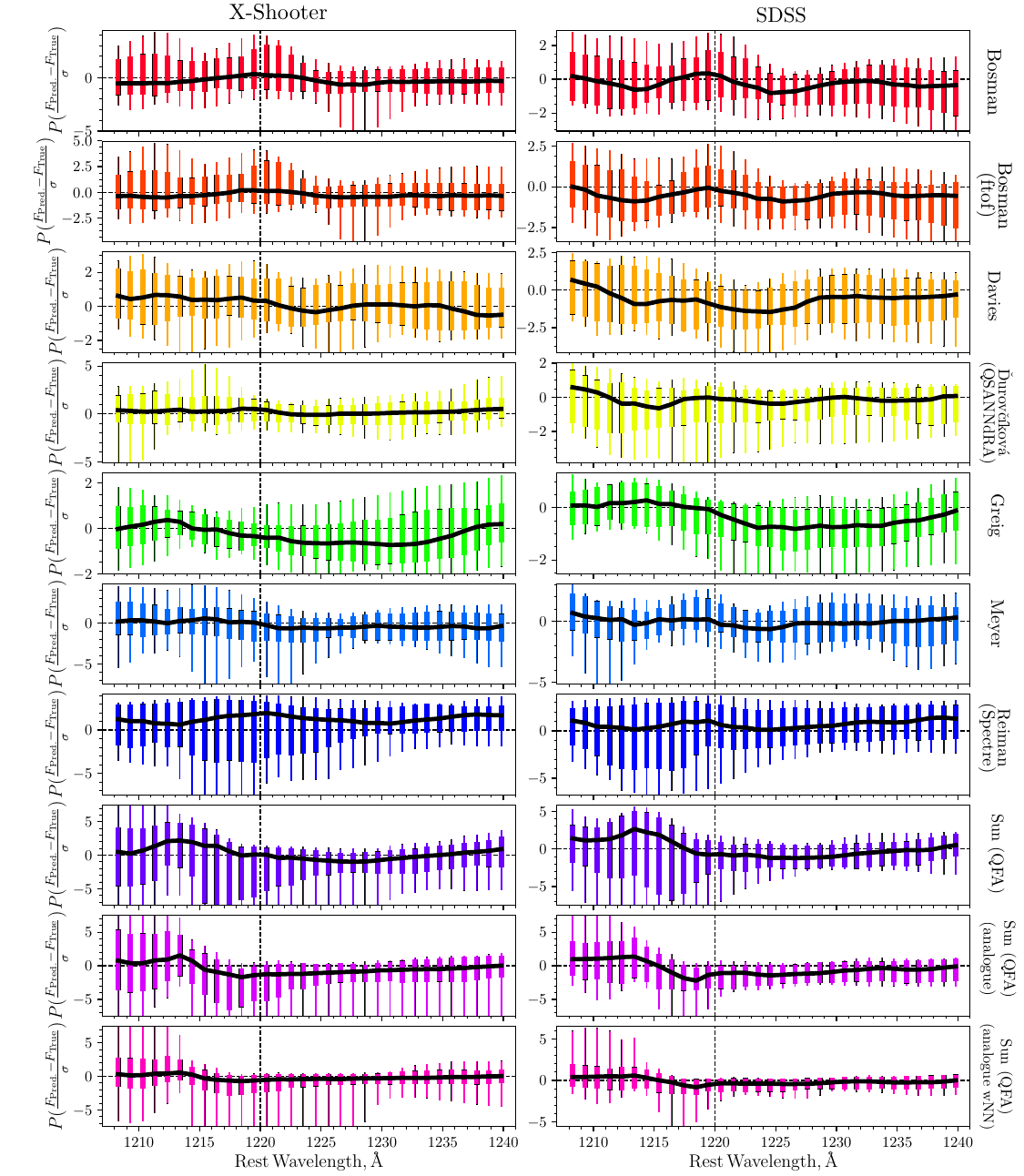}
	\end{center}
\caption[]{The PDF of predicted QSO flux relative to the true QSO flux divided by the corresponding uncertainty predicted for the QSO reconstruction pipeline. We compute this over the entire QSO sample (X-Shooter; right and SDSS; left) as a function of wavelength for each reconstruction pipeline individually. We provide this PDF for all QSO reconstruction pipelines, with the thick bars and thin lines denoting the 68th and 95th percentiles respectively while the black curve with white diamonds represent the median.}
\label{fig:perbin}
\end{figure*}

Firstly, in Figure~\ref{fig:perbin} we demonstrate the PDFs of the predicted QSO flux minus the true QSO flux weighted by the corresponding prediction uncertainty provided by each reconstruction pipeline (this is similar to Figure~\ref{fig:perbin-rel} except taking into account the prediction uncertainties). The thick bars and thin lines correspond to the 68th and 95th percentiles respectively while the black curve with white diamonds denotes the median. In representing the data in this way, we can more readily identify sample biases in over/underestimating the true QSO flux. Broadly speaking, across all pipelines the distribution of reconstructed flux tends to be skewed towards underestimating the true QSO flux, with the median sitting below zero and the tails of the distributions being more largely negatively skewed. Nevertheless, all pipelines are consistent with zero within their 68th percentile uncertainties, thus showing no systematically large biases. 

Overall, we find similar trends in performance as above, with the Greig pipeline recovering the narrowest distribution (smallest dynamic range along the vertical axis) owing to its considerably larger modelling uncertainties. Note, we again recover the same systematic offset (underestimate of the QSO flux) between [1220, 1240]~\AA. Behind the Greig pipeline, the Meyer, QSANNdRA, Davies and both Bosman pipelines have fairly similar distributions. However, both the QSANNdRA and Meyer pipelines have better overall performance highlighted by the median being closer to zero across the full wavelength range with less systematic offsets in their PDFs as a function of wavelength. For example, the Bosman and Davies pipelines have small systematic offsets (0.5-1$\sigma$), more prevalent in the SDSS sample, than those of the QSANNdRA and Meyer pipeline. For the QFA pipelines, we recover relatively broader PDFs around \lya{}, indicative of the larger scatter we observed in Figure~\ref{fig:Sample_means} owing to the tendency to over predict the QSO flux in QSOs with weak \lya{} emission (evident by the PDFs being above zero). However, as highlighted previously, redward of $1220$\AA\ QFA performs as well as the better performing pipelines.

To quantify the total performance of the reconstruction pipeline for an individual QSO, we calculate,
\begin{eqnarray} \label{eq:prob}
\Delta P_{\rm tot} = \prod^{\lambda\,{\rm range}}_{i} P(F_{\rm truth}(\lambda_{i})|{\rm model\,pipeline}),
\end{eqnarray}
which is the total probability to obtain the intrinsic (true) QSO flux ($F_{\rm truth}$) given the specific \lya{} model prediction over some spectral region, e.g. [1208, 1220]~\AA. This quantity more accurately accounts for the deviation of the predicted flux relative to the intrinsic flux whilst also properly weighting by the overall amplitude of the modelling uncertainties. In effect, it will down weight the Greig pipeline as although the relative distance away from the intrinsic flux will be lowest in each wavelength bin, the width of the distribution will ensure the overall probability is equally low. Further, it will also down weight pipelines with small uncertainties as the probability will become vanishingly small for relatively small deviations away from the intrinsic truth.

\begin{figure*} 
	\begin{center}
	  \includegraphics[trim = 0cm 0.5cm 0cm 0.5cm, scale = 0.8]{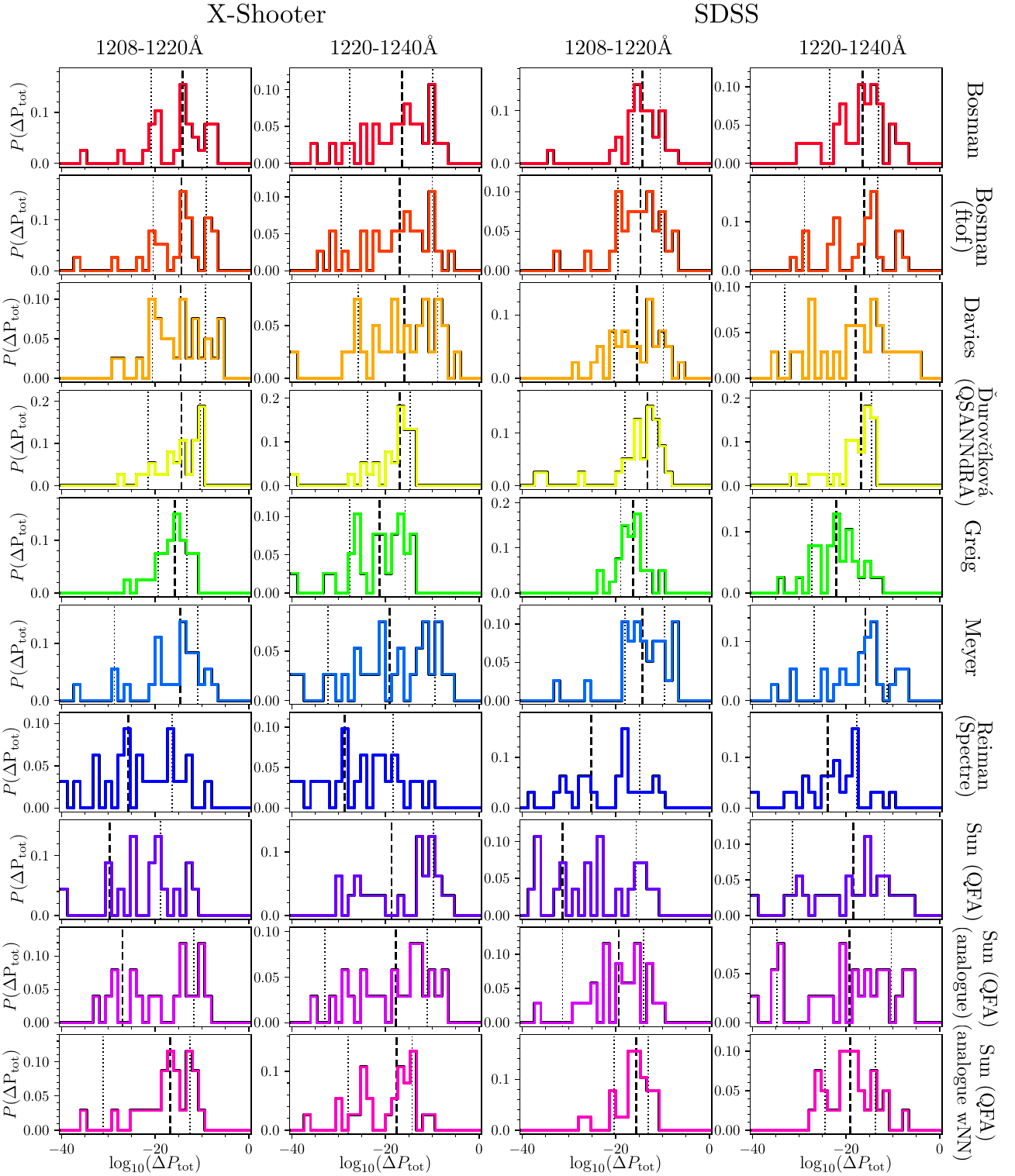}
	\end{center}
\caption[]{The PDF of ${\rm log}_{10}(\Delta P_{\rm tot})$ determined for each QSO sample (X-Shooter and SDSS) and for each wavelength coverage, [1208, 1220]~\AA\ and [1220, 1240]~\AA. The vertical dashed line corresponds to the median while the dotted lines denote the 68th percentile distribution.}
\label{fig:Ptot}
\end{figure*}

\begin{table*}
%\tiny
\caption{The median and 68th percentile uncertainties on the total probability [${\rm log}_{10}(\Delta P_{\rm tot})$] for the entire QSO sample (X-Shooter or SDSS) between our two chosen spectral regions, [1208, 1220]~\AA\ corresponding to the \lya{} peak and [1220, 1240]~\AA. The circled values indicate the best performing pipeline for each sample and spectral region.}
%\begin{adjustwidth}{-0.6cm}{}
%\begin{adjustwidth}{}{}
%\scalebox{0.8}[0.75]{
%\begin{adjustwidth}{-2cm}{}
\begin{tabular}{@{}lccccc}
\hline
Pipeline & X-Shooter & X-Shooter & SDSS & SDSS \\
Wavelength range (\AA) & [1208,1220] & [1220,1240] & [1208,1220] & [1220,1240] \\
\hline
Bosman & \circled{-14.14$\substack{+5.26\\-6.67}$} & -16.49$\substack{+6.58\\-11.15}$ & -14.25$\substack{+3.75\\-2.17}$ & -16.47$\substack{+3.37\\-6.97}$\\
Bosman (ftof) & -14.39$\substack{+5.37\\-6.03}$& -16.98$\substack{+7.00\\-12.49}$ & -14.69$\substack{+4.45\\-4.81}$ & -16.12$\substack{+2.93\\-12.77}$ \\
Davies  & -14.54$\substack{+5.35\\-6.06}$& \circled{-15.96$\substack{+7.14\\-9.78}$} & -15.49$\substack{+5.66\\-4.84}$ & -17.91$\substack{+7.10\\-15.21}$\\
 \v{D}urov\v{c}\'{i}kov\'{a} (QSANNdRA)  & -14.38$\substack{+3.92\\-7.18}$& -16.89$\substack{+2.18\\-6.97}$ & \circled{-13.24$\substack{+2.08\\-4.82}$} & -16.82$\substack{+2.33\\-6.76}$  \\
 Greig  & -15.76$\substack{+2.56\\-3.64}$& -21.22$\substack{+5.43\\-6.43}$ & -16.31$\substack{+2.85\\-2.63}$ & -22.12$\substack{+5.02\\-5.22}$\\
 Meyer  & -14.58$\substack{+3.71\\-14.13}$& -19.12$\substack{+9.68\\-13.09}$ & -14.26$\substack{+4.72\\-3.73}$ & \circled{-15.85$\substack{+4.63\\-11.02}$}\\
 Reiman (Spectre)  & -25.77$\substack{+9.44\\-16.37}$& -28.76$\substack{+10.40\\-25.55}$ & -25.26$\substack{+10.41\\-15.67}$ & -23.91$\substack{+6.24\\-19.56}$\\
 Sun (QFA)  & -29.70$\substack{+10.82\\-85.01}$& -18.67$\substack{+8.98\\-27.73}$ & -31.31$\substack{+15.70\\-40.95}$ & -18.49$\substack{+6.66\\-12.92}$\\
 Sun (QFA analogue) & -26.97$\substack{+15.26\\-55.49}$& -17.80$\substack{+6.72\\-15.10}$ & -19.34$\substack{+5.23\\-12.03}$ & -19.28$\substack{+8.95\\-15.54}$\\
 Sun (QFA analogue wNN) & -16.76$\substack{+4.15\\-14.42}$& -17.65$\substack{+3.34\\-10.29}$ & -15.63$\substack{+2.54\\-4.69}$ & -19.15$\substack{+5.46\\-5.33}$\\
\hline
\end{tabular}
%}
\label{tab:dist}
%\end{adjustwidth}
\end{table*}

In Figure~\ref{fig:Ptot} we provide the PDFs of our measured total probability, $\Delta P_{\rm tot}$ for each QSO reconstruction pipeline containing prediction uncertainties determined over the entire sample of 30 QSOs within either the X-Shooter or SDSS samples separated also by spectral coverage, [1208,1220]~\AA\ and [1220,1240]~\AA. For each of these PDFs, we also provide the median (dashed) and 68th percentile distributions which are also summarised in Table~\ref{tab:dist}. Generally speaking, the PDFs are typically broader for the X-Shooter sample than for the SDSS sample, reflecting earlier observations that the X-Shooter sample might contain more QSOs that are less consistent (i.e. outliers) with the SDSS training sets of these pipelines.

Focussing first on the \lya{} line region, the Bosman, Davies, QSANNdRA and Meyer pipelines perform the best, with the consistently lowest median ${\rm log}_{10}(\Delta P_{\rm tot})$. This, however, is not too surprising as these pipelines tend to share relatively similar methodologies, that is spectral decomposition either through PCA or in using an auto-encoder in the case of the Meyer pipeline. Interestingly, the QSANNdRA pipeline does not get significantly down weighted owing to its smaller prediction uncertainties, highlighting its strong performance at predicting the \lya{} line flux. The Greig and QFA (analogue wNN) are typically about a dex lower. For Spectre and the remaining QFA models, the discrepancy is much larger, due to the issues noted previously, the possible misuse of our implementation of Spectre and the over prediction of the \lya{} flux in QSOs exhibiting weak \lya{} line emission for QFA coupled with the considerably narrower modelling uncertainties compared to the QFA (analogue wNN) model.

Redward of \lya{}, the better performing reconstruction pipelines are again the PCA based (Bosman, Davies and QSANNdRA) or dimensionality reducing machine-learning pipelines (Meyer), although notably the Meyer pipeline is a few dex lower for the X-Shooter sample relative to the SDSS sample. About 1-2 dex behind these are both QFA analogue pipelines with the original QFA being a further dex behind those. Within this spectral region (unlike the \lya{} region) our metric, $\Delta P_{\rm tot}$, the total probability determined over a wavelength region, suitably down weights the performance of the Greig pipeline owing to its systematic offset displayed in Figure~\ref{fig:Sample_means}. Equally, QSANNdRA is generally about a dex lower than the best PCA pipelines within this region as it has a slight tendency to over predict the QSO flux coupled with its notably smaller modelling uncertainties which leads to lower values of $\Delta P_{\rm tot}$.

Overall, this solidifies our earlier observations, that reconstruction pipelines based on some form of spectral decomposition are typically the better performing pipelines. Further, that those embedded in a machine learning framework tend to perform best (Meyer and QSANNdRA), highlighting that these are better at picking up additional (non-linear) information.

%\begin{landscape}
\begin{table*}
%\tiny
\caption{The total probability [${\rm log}_{10}(\Delta P_{\rm tot})$] for obtaining the true intrinsic QSO flux of each QSO in Sample 1 (X-Shooter) for each individual QSO reconstruction pipeline comprising of modelling uncertainties (see text for further details). We consider two spectral regions, [1208, 1220]~\AA\ corresponding to the \lya{} peak and [1220, 1240]~\AA. The probability is normalised by the best performing pipeline for that spectral region and is denoted by a circle.}
\begin{adjustwidth}{-0.6cm}{}
\scalebox{0.8}[0.75]{
%\begin{adjustwidth}{-2cm}{}
\begin{tabular}{@{}lcccccccccccc}
\hline
Sample \#1 & & wavelength &  Bosman & Bosman & Davies & \v{D}urov\v{c}\'{i}kov\'{a} & Greig & Meyer & Reiman & Sun & Sun (QFA) & Sun (QFA) \\
(QSO \#) QSO name & $z$ & range (\AA) & & (ftof) & & (QSANNdRA) &  & & (Spectre) & (QFA)  & (analogue) & (analogue wNN) \\
\hline
%\vspace{0.8mm}
(1) BR J0030-5159 & 4.17 & [1208,1220] & -0.85 & \circled{0.00} & -0.48 & -3.43 & -3.32 & -18.26 & -25.46 & -150.17 & -65.48 & -7.13 \\
 & & [1220,1240] & -34.98 & -20.20 & -7.32 & -2.64 & -0.07 & -11.39 & -4.06 & -118.01 & -21.08 & \circled{0.00}\\
\hline
(2) HB89 0055-269  & 3.66 & [1208,1220] &  -0.93 & -1.11 & \circled{0.00} & -3.96 & -5.05 & -0.43 & -12.02 & -7.58 & -0.17 & -1.87 \\
 & & [1220,1240]& -10.61 & -10.53 & -1.27 & \circled{0.00} & -9.76 & -3.28 & -11.49 & -5.61 & -11.06 & -8.54 \\
\hline
(3) PMN J0100-2708 & 3.55 &[1208,1220] & -2.35 & -2.70 & \circled{0.00} & -4.49 & -6.25 & -14.28 & -11.97 & -9.74 & -4.08 & -7.10 \\
 & & [1220,1240]& -2.84 & -3.00 & \circled{0.00} & -7.63 & -11.07 & -1.44 & -11.88 & -2.65 & -6.32 & -7.41 \\
\hline
(4) PSS J0117+1552 &4.24 &[1208,1220] &-1.06 & -1.12 & -1.42 & -7.11 & -5.64 & \circled{0.00} & -3.41 & -13.82 & -15.63 & -5.85 \\
 & & [1220,1240]& -1.05 & -0.69 & \circled{0.00} & -1.16 & -4.90 & -3.58 & -4.56 & -4.32 & -1.34 & -3.40\\
 \hline
(5) BR J0401-1711 &4.23 &[1208,1220] & -17.76 & -18.16 & \circled{0.00} & -2.95 & -1.84 & -8.38 & -174.95 & -849.65 & -399.65 & -801.91\\
 & & [1220,1240]& -33.24 & -37.53 & -5.40 & \circled{0.00} & -0.34 & -22.14 & -92.01 & -242.83 & -207.44 & -151.20 \\
 \hline
(6) BR J0529-3526 & 4.42&[1208,1220] & -0.34 & \circled{0.00} & -2.20 & -4.73 & -4.84 & -1.18 & -1.63 & -5.62 & -21.90 & -3.92\\
 & & [1220,1240]& -1.96 & -1.41 & \circled{0.00} & -9.60 & -9.15 & -6.16 & -7.46 & -1.55 & -8.73 & -6.33 \\
\hline
(7) SDSS J075552.41 & 3.67&[1208,1220] & -1.24 & -1.25 & \circled{0.00} & -3.07 & -6.71 & -0.37 & -16.41 & -22.67 & -7.82 & -4.85\\
 \qquad \qquad+134551.1 & &[1220,1240] & -3.52 & -2.94 & -2.10 & -10.09 & -10.44 & \circled{0.00} & -9.63 & -0.29 & -2.81 & -7.67 \\
\hline
(8) SDSS J083322.50& 3.72&[1208,1220] & -0.06 & \circled{0.00} & -1.49 & -0.98 & -2.98 & -4.40 & -23.92 & -8.80 & -1.18 & -3.48\\
 \qquad \qquad+095941.2 & &[1220,1240] & -3.28 & -3.15 & \circled{0.00} & -5.82 & -6.40 & -0.75 & -43.60 & -0.64 & -0.98 & -6.25 \\
\hline
(9) SDSS J093714.48 &3.70 &[1208,1220] & -16.51 & -17.41 & -1.60 & -1.66 & -4.70 & \circled{0.00} & -1.40 & -31.10 & -38.29 & -11.71 \\
 \qquad \qquad+082858.6 & &[1220,1240] & -4.00 & -4.46 & \circled{0.00} & -6.00 & -5.38 & -1.13 & -7.89 & -11.98 & -5.72 & -6.66\\
\hline
(10) J101818+054822 & 3.52&[1208,1220] & -7.38 & -6.19 & -6.10 & -6.08 & -3.54 & -14.71 & \circled{0.00} & -5.93 & -11.49 & -2.70 \\
 & &[1220,1240] & -9.99 & -10.28 & -13.63 & -7.47 & -9.15 & -5.10 & -13.18 & -1.18 & \circled{0.00} & -2.51 \\
\hline
(11) BR 1033-0327 &4.53 &[1208,1220] & -0.02 & \circled{0.00} & -0.86 & -1.65 & -4.39 & -4.05 & -8.07 & -4.43 & -1.34 & -1.37 \\
 & &[1220,1240] &  -2.71 & -1.90 & -4.82 & -5.85 & -8.13 & \circled{0.00} & -13.75 & -2.52 & -1.25 & -6.52  \\
\hline
(12) SDSSJ1037+2135& 3.63&[1208,1220] & -5.93 & -6.48 & -3.17 & \circled{0.00} & -5.72 & -2.78 & -3.42 & -10.89 & -6.37 & -3.94 \\
 & &[1220,1240] & -3.78 & -2.83 & -2.77 & -3.02 & -4.33 & -2.46 & \circled{0.00} & -20.08 & -7.96 & -5.65 \\
\hline
(13) SDSSJ1042+1957&3.63 &[1208,1220] & -1.21 & -1.00 & -2.94 & -5.83 & \circled{0.00} & -0.93 & -10.06 & -95.94 & -74.40 & -17.60 \\
 & &[1220,1240] & -7.87 & -8.11 & -2.59 & -5.49 & -11.55 & -2.48 & -18.88 & \circled{0.00} & -1.99 & -6.66 \\
\hline
(14) SDSS J105340.75& 3.66& [1208,1220]&- 0.81 & -0.86 & \circled{0.00} & -4.78 & -7.73 & -4.65 & -32.91 & -18.06 & -2.30 & -6.02 \\
 \qquad \qquad+010335.6 & &[1220,1240] & -2.76 & -3.06 & \circled{0.00} & -9.96 & -7.32 & -0.46 & -55.30 & -3.46 & -1.14 & -6.66 \\
\hline
(15) J111701+131115& 3.62&[1208,1220] &-6.50 & -7.37 & -14.00 & -0.90 & \circled{0.00} & -6.36 & -7.52 & -11.57 & -10.13 & -1.94 \\
 & &[1220,1240] & -4.56 & -4.74 & -3.69 & -5.77 & -8.55 & -0.09 & -27.09 & \circled{0.00} & -3.45 & -9.09 \\
\hline
(16) J112617-012632& 3.63&[1208,1220] & \circled{0.00} & -0.28 & -2.97 & -2.99 & -6.61 & -0.34 & -13.92 & -107.81 & -92.58 & -141.74 \\
 & &[1220,1240] & -0.00 & \circled{0.00} & -6.76 & -8.96 & -13.12 & -25.17 & -5.80 & -21.77 & -9.54 & -7.31 \\
\hline
(17) HB89 1159+123& 3.52&[1208,1220] &-1.73 & -1.12 & -0.04 & -31.17 & -4.29 & \circled{0.00} & -7.29 & -43.61 & -38.72 & -9.74 \\
 & &[1220,1240] & \circled{0.00} & -0.35 & -1.41 & -2.99 & -7.95 & -4.11 & -9.58 & -5.84 & -2.43 & -5.26 \\
\hline
(18) SDSSJ1202-0054& 3.59&[1208,1220] &-11.22 & -11.85 & -8.16 & -4.06 & -3.56 & \circled{0.00} & -46.84 & -93.43 & -191.55 & -52.94 \\
 & &[1220,1240] & -1.85 & -2.26 & -1.52 & -2.35 & \circled{0.00} & -2.99 & -6.83 & -9.71 & -9.27 & -14.64 \\
\hline
(19) J124837+130440&3.72 &[1208,1220] & -0.04 & \circled{0.00} & -8.36 & -2.43 & -3.28 & -23.68 & -3.09 & -6.82 & -2.31 & -1.36 \\
 & &[1220,1240] & -4.21 & -3.76 & -6.57 & -4.02 & -15.91 & -8.31 & -8.78 & \circled{0.00} & -0.67 & -4.90 \\
\hline
(20) J135247+130311& 3.71&[1208,1220] &-4.26 & -4.11 & \circled{0.00} & -0.49 & -5.53 & -5.01 & -16.82 & -10.12 & -2.80 & -3.27 \\
 & &[1220,1240] & -1.98 & -1.29 & \circled{0.00} & -1.27 & -7.92 & -24.41 & -15.43 & -0.02 & -1.29 & -3.33 \\
\hline
(21) SDSSJ1416+1811& 3.59&[1208,1220] & -4.35 & -4.22 & -13.10 & -5.96 & \circled{0.00} & -40.55 & -19.70 & -39.26 & -55.38 & -13.30 \\
 & &[1220,1240] & -1.60 & -1.56 & -3.60 & -9.23 & -10.20 & -2.71 & -2.60 & -0.80 & \circled{0.00} & -2.03 \\
\hline
(22) PKS B1418-064 & 3.69&[1208,1220] & -3.42 & -3.71 & -3.69 & \circled{0.00} & -4.39 & -10.61 & -8.72 & -0.50 & -13.85 & -5.54 \\
 & &[1220,1240] & 2.71 & -2.63 & \circled{0.00} & -5.04 & -8.30 & -3.21 & -27.02 & -1.73 & -3.62 & -13.53 \\
\hline
(23) J144250+092001 & 3.53&[1208,1220] & -1.09 & -1.00 & -2.59 & \circled{0.00} & -4.94 & -1.26 & -31.80 & -36.01 & -9.15 & -6.27 \\
 & &[1220,1240] & -3.85 & -4.08 & -0.41 & \circled{0.00} & -10.68 & -1.06 & -23.76 & -1.95 & -7.68 & -8.62 \\
\hline
(24) SDSSJ1445+0958 &3.56 &[1208,1220] & -0.19 & \circled{0.00} & -3.57 & -9.74 & -3.71 & -0.71 & -28.63 & -17.10 & -0.77 & -3.61 \\
 & &[1220,1240] & -0.22 & \circled{0.00} & -10.53 & -6.59 & -4.82 & -3.94 & -49.75 & -2.83 & -0.48 & -5.71 \\
\hline
(25) J150328+041949& 3.69&[1208,1220] & -1.96 & -2.22 & \circled{0.00} & -117.82 & -2.00 & -28.55 & -14.38 & -62.52 & -43.28 & -6.97 \\
 & &[1220,1240] & -2.54 & -1.69 & \circled{0.00} & -5.32 & -6.14 & -13.21 & -38.60 & -7.15 & -6.02 & -6.09 \\
\hline
(26) SDSSJ1517+0511& 3.55&[1208,1220] & -5.08 & -6.10 & \circled{0.00} & -3.91 & -9.24 & -2.92 & -21.29 & -14.45 & -4.75 & -5.34 \\
 & &[1220,1240] & -6.79 & -8.36 & \circled{0.00} & -6.01 & -9.11 & -4.08 & -15.72 & -3.87 & -12.43 & -9.62 \\
\hline
(27) J1621-0042& 3.71&[1208,1220] & -4.26 & -3.47 & -5.87 & -1.21 & -7.02 & \circled{0.00} & -29.67 & -114.40 & -38.71 & -5.88 \\
 & &[1220,1240] & -6.28 & -6.24 & -14.43 & -20.61 & -1.35 & -14.20 & -7.73 & -29.30 & -15.54 & \circled{0.00} \\
\hline
(28) PSS J1723+2243&4.53 &[1208,1220] & -1.74 & -1.99 & -7.85 & -1.89 & -3.74 & -6.41 & \circled{0.00} & -26.34 & -20.66 & -3.24 \\
 & &[1220,1240] & -10.05 & -5.87 & -0.85 & -14.13 & -6.37 & \circled{0.00} & -35.18 & -2.05 & -163.12 & -11.25 \\
\hline
(29) J2239536-055219& 4.56&[1208,1220] & -1.84 & -2.21 & -1.56 & -4.99 & -6.64 & \circled{0.00} & -10.50 & -28.00 & -18.57 & -5.01 \\
 & &[1220,1240] &  -0.09 & \circled{0.00} & -3.12 & -4.33 & -24.20 & -5.52 & -14.67 & -41.09 & -16.36 & -3.68  \\
\hline
(30) BR 2248-1242& 4.16&[1208,1220] & -27.71 & -54.35 & -2.49 & -0.50 & \circled{0.00} & -201.06 & -739.60 & - & - & - \\
 & &[1220,1240] & -8.04 & -26.68 & -3.48 & \circled{0.00} & -31.93 & -166.32 & -24.97 & -98.66 & - & - \\
\hline
%\end{adjustwidth}
\end{tabular}}
%\footnotesize{\\$^a$ Although identified as a possible mini/weak BAL by \citet{Bischetti:2023}, the associated BAL complex does not impact the \lya{}-\nv{} emission line complex. $^b$ a potential weak BAL complex although \citet{Becker:2015} deem it more likely to be an intrinsic feature of the QSO.}
\label{tab:Sample1}
\end{adjustwidth}
\end{table*}
%\end{landscape}

%\begin{landscape}
\begin{table*}
%\tiny
\caption{The same as Table~\ref{tab:Sample1} except for Sample 2 (SDSS) spectra.}
\begin{adjustwidth}{-0.6cm}{}
\scalebox{0.8}[0.75]{
%\begin{adjustwidth}{-2cm}{}
\begin{tabular}{@{}lcccccccccccc}
\hline
Sample \#2 & & wavelength &  Bosman & Bosman & Davies & \v{D}urov\v{c}\'{i}kov\'{a} & Greig & Meyer & Reiman & Sun & Sun (QFA) & Sun (QFA) \\
(QSO \#) QSO name & $z$ & range (\AA) & & (ftof) & & (QSANNdRA) & & & (Spectre) &  & (analogue) & (analogue wNN) \\
\hline
%\vspace{0.8mm}
(1) 3682-55244-942 & 3.76 & [1208,1220] & -4.51 & -3.12 & -8.54 & -7.43 & -3.03 & \circled{0.00} & -0.84 & -2.86 & -2.78 & -0.96 \\
 & & [1220,1240] &-4.18 & -5.47 & -8.29 & -8.25 & -12.46 & \circled{0.00} & -3.50 & -14.96 & -24.06 & -7.93 \\
\hline
(2) 3776-55209-946  & 3.60 & [1208,1220] & -19.36 & -17.50 & -13.93 & \circled{0.00} & -1.20 & -3.06 & -67.67 & -132.01 & -3.82 & -0.94 \\
 & & [1220,1240]& -10.70 & -8.64 & -10.50 & -11.65 & -10.16 & -12.45 & -26.56 & -36.41 & \circled{0.00} & -0.67 \\
\hline
(3) 3779-55222-358 & 3.63 &[1208,1220] & -2.78 & -2.31 & -9.89 & -2.36 & -0.30 & -2.31 & -4.22 & -50.11 & \circled{0.00} & -4.10 \\
 & & [1220,1240]& -11.11 & -14.56 & -18.45 & -15.18 & -17.29 & \circled{0.00} & -27.51 & -1.44 & -17.03 & -3.91 \\
\hline
(4) 3839-55575-164 &3.63 &[1208,1220] & -0.31 & \circled{0.00} & -0.37 & -1.07 & -3.83 & -4.27 & -5.57 & -5.23 & -7.22 & -1.77 \\
 & & [1220,1240]& -0.73 & \circled{0.00} & -18.16 & -2.39 & -8.50 & -21.09 & -3.75 & -8.19 & -3.70 & -4.20 \\
 \hline
(5) 3860-55269-816 &3.63 &[1208,1220] & -1.34 & -1.24 & \circled{0.00} & -1.98 & -8.16 & -0.80 & -7.82 & -19.56 & -1.84 & -5.28\\
 & & [1220,1240]& \circled{0.00} & -11.40 & -29.09 & -8.12 & -4.83 & -6.82 & -0.83 & -8.49 & -23.21 & -3.79 \\
 \hline
(6) 3872-55382-651 & 3.65&[1208,1220] & -2.29 & -2.11 & -6.26 & -3.13 & -5.30 & \circled{0.00} & -7.45 & -14.85 & -16.96 & -5.71\\
 & & [1220,1240]& -3.02 & -2.39 & -3.75 & -8.22 & -6.72 & -0.92 & -22.29 & \circled{0.00} & -1.31 & -7.57 \\
\hline
(7) 4011-55635-724 & 3.60&[1208,1220] & \circled{0.00} & -10.12 & -3.40 & -3.08 & -2.08 & -1.56 & -10.07 & -39.35 & -20.22 & -9.41\\
  & &[1220,1240] & -2.69 & -2.30 & \circled{0.00} & -2.97 & -7.72 & -3.10 & -15.54 & -0.25 & -2.43 & -5.57 \\
\hline
(8) 4178-55653-342& 3.77&[1208,1220] & -4.09 & -6.45 & -3.50 & \circled{0.00} & -5.55 & -20.10 & -4.07 & -13.35 & -9.10 & -1.24\\
  & &[1220,1240] & -0.72 & -0.35 & -1.56 & -3.23 & -8.21 & -2.90 & -8.98 & \circled{0.00} & -0.75 & -5.63 \\
\hline
(9) 4229-55501-552 &3.84 &[1208,1220] & -0.25 & -1.33 & -2.20 & -2.57 & -1.68 & \circled{0.00} & -1.99 & -9.82 & -1.52 & -2.19 \\
 & &[1220,1240] &-5.22 & -7.70 & -31.03 & -1.27 & -8.78 & -10.33 & \circled{0.00} & -2.70 & -18.92 & -5.06 \\
\hline
(10) 4361-55831-656 & 3.52&[1208,1220] & -2.39 & \circled{0.00} & -1.88 & -2.70 & -2.87 & -0.64 & -23.99 & -30.04 & -10.75 & -5.33 \\
 & &[1220,1240] & -3.44 & -1.70 & \circled{0.00} & -6.89 & -11.17 & -1.60 & -32.59 & -1.78 & -0.25 & -6.18 \\
\hline
(11) 4629-55630-210 &3.95 &[1208,1220] & -0.81 & -5.41 & \circled{0.00} & -1.37 & -3.58 & -2.82 & -40.46 & -147.39 & -8.82 & -1.56 \\
 & &[1220,1240] &-1.11 & -6.13 & -9.88 & -1.12 & -3.96 & -13.01 & \circled{0.00} & -21.67 & -7.98 & -8.30 \\
\hline
(12) 4711-55737-766 &3.62 &[1208,1220] &-0.50 & -0.32 & -1.13 & -0.23 & -3.13 & -0.50 & -0.94 & -2.53 & -0.30 & \circled{0.00} \\
 & &[1220,1240] & -3.48 & -4.43 & -11.28 & -3.65 & -13.11 & \circled{0.00} & -10.45 & -2.32 & -14.16 & -5.30 \\
\hline
(13) 4747-55652-761& 3.66&[1208,1220] & -3.40 & -6.32 & -2.36 & -1.05 & -4.71 & \circled{0.00} & -0.22 & -3.17 & -11.67 & -2.61 \\
 & &[1220,1240] &-6.71 & -9.52 & -12.82 & -4.44 & -13.67 & \circled{0.00} & -3.24 & -15.85 & -21.26 & -3.19 \\
\hline
(14) 4799-55656-446&4.11 &[1208,1220] & -2.95 & -2.81 & -9.17 & -1.36 & -5.33 & \circled{0.00} & -14.74 & -20.53 & -4.99 & -4.16\\
 & &[1220,1240] & -19.70 & -29.00 & -24.03 & -36.84 & -3.23 & -18.77 & \circled{0.00} & -5.61 & -17.04 & -1.65\\
\hline
(15) 5174-56047-750& 3.55& [1208,1220]&-1.64 & -6.14 & -7.05 & \circled{0.00} & -2.82 & -33.56 & -6.12 & -2.33 & -1.18 & -1.92\\
 & &[1220,1240] &-1.56 & -0.79 & -0.86 & \circled{0.00} & -6.56 & -3.24 & -50.42 & -0.96 & -1.74 & -5.95 \\
\hline
(16) 5184-56352-692& 3.64&[1208,1220] &-1.80 & -2.64 & -3.12 & \circled{0.00} & -5.96 & -1.23 & -15.96 & -9.60 & -4.83 & -3.23 \\
 & &[1220,1240] & -4.90 & -5.06 & -5.04 & \circled{0.00} & -6.54 & -1.82 & -1.80 & -2.38 & -1.80 & -3.94 \\
\hline
(17) 5376-55987-382& 3.69&[1208,1220] & -0.91 & -1.30 & \circled{0.00} & -0.05 & -6.03 & -1.87 & -18.08 & -21.39 & -6.23 & -5.22 \\
 & &[1220,1240] & \circled{0.00} & -0.15 & -0.01 & -0.68 & -5.96 & -6.98 & -32.59 & -2.06 & -2.14 & -7.42 \\
\hline
(18) 5443-56010-898& 3.69&[1208,1220] &-1.54 & -1.66 & -1.15 & -0.10 & -7.68 & \circled{0.00} & -24.13 & -24.48 & -6.09 & -7.74 \\
 & &[1220,1240] & -1.50 & -2.82 & -1.12 & -1.27 & -9.11 & -1.81 & -14.30 & \circled{0.00} & -2.37 & -6.27 \\
\hline
(19) 5444-56038-798& 3.68&[1208,1220] &-2.08 & \circled{0.00} & -3.45 & -2.35 & -5.77 & -0.34 & -0.81 & -5.59 & -11.90 & -5.17 \\
 & &[1220,1240] & -4.06 & -5.29 & \circled{0.00} & -6.46 & -14.51 & -7.29 & -1.43 & -3.71 & -3.37 & -5.12 \\
\hline
(20) 5472-55976-822&3.54 &[1208,1220] & -2.03 & -2.84 & \circled{0.00} & -15.52 & -7.13 & -4.43 & -26.22 & -26.46 & -4.72 & -7.56 \\
 & &[1220,1240] & -1.76 & -2.52 & \circled{0.00} & -8.11 & -8.43 & -1.22 & -24.26 & -0.86 & -4.54 & -7.17 \\
\hline
(21) 5715-56657-890& 3.97&[1208,1220] &-1.44 & -5.54 & -9.44 & -21.59 & -9.74 & \circled{0.00} & -14.11 & -41.35 & -35.16 & -6.80 \\
 & &[1220,1240] & \circled{0.00} & -0.40 & -6.16 & -3.70 & -38.01 & -2.73 & -2.46 & -18.73 & -13.65 & -3.28 \\
\hline
(22) 6418-56354-249& 3.77&[1208,1220] & -2.49 & -4.64 & -6.84 & -3.46 & -3.73 & \circled{0.00} & -17.48 & -73.07 & -58.14 & -13.71\\
 & &[1220,1240] &-4.88 & -4.54 & -1.12 & -4.43 & -7.48 & -4.29 & -2.97 & -0.74 & \circled{0.00} & -2.51 \\
\hline
(23) 6512-56567-682 & 3.93&[1208,1220] &-0.99 & -0.35 & -1.33 & -7.00 & -2.36 & \circled{0.00} & -33.44 & -28.59 & -19.17 & -5.07 \\
 & &[1220,1240] &-0.98 & -0.90 & -0.12 & -8.38 & -7.27 & -1.05 & -38.54 & -3.39 & \circled{0.00} & -9.37 \\
\hline
(24) 6713-56402-346 & 3.57&[1208,1220] & -1.43 & -1.63 & -4.71 & \circled{0.00} & -2.62 & -4.29 & -7.06 & -10.99 & -13.26 & -3.53 \\
 & &[1220,1240] & -2.94 & -2.56 & -2.88 & -0.63 & -8.33 & \circled{0.00} & -5.85 & -1.42 & -7.17 & -6.97 \\
\hline
(25) 6745-56425-330 &3.83 &[1208,1220] & -0.97 & -1.10 & \circled{0.00} & -3.21 & -6.51 & -0.36 & -18.47 & -13.55 & -3.09 & -5.36 \\
 & &[1220,1240] & -5.02 & -3.84 & \circled{0.00} & -9.97 & -12.34 & -3.78 & -18.88 & -1.77 & -3.75 & -6.69 \\
\hline
(26) 6758-56415-369& 3.90&[1208,1220] & -1.59 & -2.00 & -0.19 & \circled{0.00} & -9.35 & -4.19 & -7.28 & -25.55 & -10.69 & -5.90 \\
 & &[1220,1240] &-9.89 & -9.10 & -20.53 & \circled{0.00} & -6.39 & -2.35 & -5.34 & -10.29 & -14.87 & -3.59 \\
\hline
(27) 7085-56625-895& 3.77&[1208,1220] &\circled{0.00} & -0.75 & -5.10 & -17.62 & -4.23 & -7.28 & -21.74 & -43.43 & -9.31 & -1.16 \\
 & &[1220,1240] &-1.46 & -2.62 & -10.95 & \circled{0.00} & -8.35 & -25.09 & -12.29 & -11.68 & -5.04 & -8.57 \\
\hline
(28) 7096-56683-885& 3.69&[1208,1220] & -3.57 & -2.21 & \circled{0.00} & -0.17 & -1.94 & -2.61 & -53.30 & -93.08 & -25.14 & -4.91 \\
 & &[1220,1240] & -3.29 & -2.44 & -6.64 & -4.11 & -4.35 & -3.24 & -10.25 & -5.22 & -0.08 & \circled{0.00} \\
\hline
(29) 7112-56666-380&3.52 &[1208,1220] & -9.27 & -2.99 & -6.16 & -3.62 & -6.24 & \circled{0.00} & -82.55 & -176.89 & -312.37 & -226.05 \\
 & &[1220,1240] & -2.60 & -2.46 & -1.26 & -5.57 & -3.58 & \circled{0.00} & -34.05 & -43.75 & -7.88 & -4.84 \\
\hline
(30) 9616-58135-682& 4.02&[1208,1220] & -0.50 & -0.97 & -0.74 & -0.44 & -3.84 & -3.20 & -2.63 & \circled{0.00} & -7.96 & -1.66 \\
 & &[1220,1240] &-13.29 & -27.75 & -4.14 & -2.76 & -3.80 & \circled{0.00} & -4.19 & -17.27 & -6.21 & -4.52\\
\hline
%\end{adjustwidth}
\end{tabular}}
%\footnotesize{\\$^a$ Although identified as a possible mini/weak BAL by \citet{Bischetti:2023}, the associated BAL complex does not impact the \lya{}-\nv{} emission line complex. $^b$ a potential weak BAL complex although \citet{Becker:2015} deem it more likely to be an intrinsic feature of the QSO.}
\label{tab:Sample2}
\end{adjustwidth}
\end{table*}
%\end{landscape}

For completeness, in Tables~\ref{tab:Sample1} and~\ref{tab:Sample2} we summarise the total probability to obtain the intrinsic QSO flux for each individual QSO for the X-Shooter and SDSS sample, respectively. Once again, we also split the performance by wavelength range, [1208, 1220]~\AA\ and [1220, 1240]~\AA. To ease the interpretation of the data we demarcate the best performing pipeline in each individual scenario by a circle and normalise the probabilities by the pipeline with the highest probability. Further, this quantity equally acts to more readily identify problematic QSOs for each pipelines, with significantly larger values indicative of lower performance. Given the number of QSOs reconstructed across the two samples there is a large amount of information embedded in these tables.

\section{Discussions and Conclusions} \label{sec:conclusions}

We have performed a comprehensive blind study of all the available QSO \lya{} reconstruction pipelines in the literature. For this, we generated two samples of 30 QSOs observed between $3.5 < z < 4.5$ from X-Shooter and SDSS (BOSS) with their spectral information below rest-frame 1260\AA\ removed. Our blind QSOs consist of randomly selected objects exhibiting a mix of both prominent and weak \lya{} lines. This was adopted to explore the applicability of these reconstruction pipelines over a diverse population of QSOs. Ultimately, the goal of this comparison study was not to champion a single, best performing reconstruction pipeline. Rather, its goal was to explore their general performance across a broad range of QSOs while also testing several common assumptions.

In total, we consider 13 different QSO reconstruction pipelines. Broadly, these cover four different underlying methodologies to reconstruct the \lya{} line flux from the available redward ($>1260$\AA) spectral information. These include a covariance matrix of emission line correlations \citep{Greig:2017a}, PCA spectral decomposition \citep[][]{Davies:2018,Dominika:2020,Reiman:2020,Bosman:2021}, spectral decomposition through auto-encoders \citep[][Meyer et al., in prep.]{Liu:2021}, direct wavelength bin to bin flux correlations \citep{Fathivavsari:2020} and latent factor analysis \citep{Sun:2022}. Further, these employ a broad range of analytic methods or various machine learning tools to connect the information red and blueward of \lya{}.

To gain insights regarding the performance of these pipelines, we considered the following analyses. Firstly, we provided example reconstructions of individual QSOs exhibiting prominent and weak \lya{} emission to infer initial qualitative trends before also demonstrating a couple of failures, highlighting the variability of the QSOs in the sample. Next, we explored the blind sample as a whole, using our initial qualitative inferences to provide more robust interpretations. For this we considered the ratio of predicted flux to intrinsic flux for each QSO along with the mean over the entire sample. This allowed inferences to be made regarding overall pipeline performance including any observations of systematic biases. Finally, we performed a more quantitative analysis, determining the distribution of reconstructed QSOs within some fractional threshold of the true intrinsic flux. We considered this for all pipelines and then for only those that provide statistical uncertainties with their predictions. For those pipelines that provide model uncertainties we also constructed the total probability for a pipeline to obtain the intrinsic flux. Below, we summarise the main observations from these.

By considering an X-Shooter and SDSS (BOSS) sample of QSOs, we could test whether the quality of the QSO spectra in the training set impacted the performance of the reconstructions. All but one of the pipelines utilise SDSS for their training sets, with iQNet (HST) considering a completely unique sample of HST COS spectra at $z<1$. In general, we found that SDSS trained pipelines performed equally well on either the SDSS or X-Shooter sample, recovering no systematic trends or biases on average over the entire sample. When considering the reconstructed QSO distribution within a flux threshold, we did observe a slightly higher flux threshold was required to contain the same number of QSOs for the X-Shooter sample, along with an increased scatter. However, these differences were typically $\lesssim5-10$ per cent, which constitutes 1-2 QSOs. In our case, the X-Shooter sample was 4-5 times higher resolution than the SDSS (BOSS) training set. Therefore, provided the observational characteristics of the training set are relatively comparable to the target object, there does not appear to be a systematic bias.

However, the reconstructions from iQNet (HST) exhibit systematically large, but equivalent differences for both the X-Shooter and SDSS samples. For prominent \lya{} line QSOs, it tends to predict slightly narrower but more peaked \lya{} amplitudes which taper off more quickly redward of \lya{}. Thus, the higher resolution training set of HST compared to X-Shooter or SDSS may be producing quantifiable differences due to the instrument characteristics. However, the largest difference stems entirely from the fact the HST training set is dominated by QSOs with prominent \lya{} peaks. This could be due to either instrumental resolution and/or the much lower redshift of these QSOs. For QSOs with weak \lya{} emission, iQNet (HST) predicts a sharply peaked \lya{} line profile considerably overestimating the flux within that region. Therefore, the strongest source of bias is actually whether the QSO training set contains a sufficient sample of QSOs representative of the object that is to be reconstructed. This is the most important facet for the construction of the training sets. Indeed, updated QFA models used in this comparison rely on small training sets of analogue QSOs rather than large, population wide datasets to boost the overall performance of their \lya{} profile reconstructions.

Another common assumption we consider in applying our reconstruction pipelines is that QSO properties do not evolve with either redshift or luminosity. We performed an exploration of this, considering nearest neighbour reconstructions of our blind QSO sample using a large training set of QSOs separated into narrow redshift bins (at fixed luminosity) or luminosity bins (at fixed redshift). In both cases, we observed very weak trends of increasing predicted flux with increasing redshift or luminosity, however, these are entirely consistent with no evolution within the statistical uncertainties of the samples. Thus, we conclude that evolution with redshift or luminosity does not play a significant role in biasing the \lya{} predictions. Note, we do observe that the evolution is stronger for QSOs with more prominent \lya{} profiles than weaker ones (see e.g. \citealt{Dietrich:2002}), indicating that this could contribute to larger scatter in the predicted \lya{} profile amplitude. However, since the redward information tends to correlate similarly with luminosity, the predictions should not be strongly impacted. Nonetheless, we recommend that this trend be further explored with a substantially larger training set spanning a broader range of both redshifts and luminosities. We additionally explored the impact of redshift evolution by comparing the mean predictions from individual pipelines that are trained using datasets of QSOs with narrow redshift ranges compared to those with broad redshift ranges. Those with narrow ranges, $\Delta z \gtrsim 1.5$ smaller than the blind QSOs did not appear to exhibit any systematic offset, further lending weight to our assumptions of no significant redshift evolution.

Overall, we found the better performing pipelines to consistently rely on machine learning approaches. This is not too surprising as generally speaking machine learning is designed to better pick up non-linear and more complex features in the properties of the large datasets or of complex spectra. However, one must be extremely careful with such approaches as in some cases these only provide a singular best-guess with no statistical uncertainties. Others may only partially characterise the uncertainties, considering some but not all of the modelling uncertainties and thus underestimate the true distribution of scatter from the underlying QSO population. Secondly, \lya{} reconstruction pipelines that rely on some form of spectral decomposition, for example principal component analysis, auto-encoders or analogously factor analysis, generally perform best. Again, these rely on drastically reducing the wealth of information into a series of spectral components governing the correlations and a set of variable amplitudes.

Finally, the relative performance described above is based on the same unified set of blind QSOs for vastly different methodologies learnt on different training sets. To truly ascertain the best performing approach for \lya{} profile reconstruction, all different methodologies would need to be trained with the same training set. Such an approach is beyond the scope of this current comparison, and thus we defer it to the future.

\section*{Acknowledgements}

We thank John Tamanas who provided initial assistance in setting up and running Spectre prior to leaving academia. Parts of this research were supported by the Australian Research Council Centre of Excellence for All Sky Astrophysics in 3 Dimensions (ASTRO 3D), through project number CE170100013. S.E.I.B. is supported by the Deutsche Forschungsgemeinschaft (DFG) under Emmy Noether grant number BO 5771/1-1. R.A.M acknowledges support from the Swiss National Science Foundation (SNSF) through project grant 200020\_207349. A.M. acknowledges support from the Ministry of Universities and Research (MUR) through the PRIN project ”Optimal inference from radio images of the epoch of reionization” as well as the PNRR project ”Centro Nazionale di Ricerca in High Performance Computing, Big Data e Quantum Computing”. Y.S.T. acknowledges financial support from the Australian Research Council through DECRA Fellowship DE220101520. This work was performed on the OzSTAR national facility at Swinburne University of Technology. The OzSTAR program receives funding in part from the Astronomy National Collaborative Research Infrastructure Strategy (NCRIS) allocation provided by the Australian Government, and from the Victorian Higher Education State Investment Fund (VHESIF) provided by the Victorian Government. The XQ-100 sample is based on observations made with ESO Telescopes at the La Silla Paranal Observatory under programme ID 189.A-0424.

Funding for the Sloan Digital Sky Survey IV has been provided by the Alfred P. Sloan Foundation, the U.S. Department of Energy Office of Science, and the Participating Institutions. SDSS acknowledges support and resources from the Center for High-Performance Computing at the University of Utah. The SDSS web site is \url{www.sdss4.org}.

SDSS is managed by the Astrophysical Research Consortium for the Participating Institutions of the SDSS Collaboration including the Brazilian Participation Group, the Carnegie Institution for Science, Carnegie Mellon University, Center for Astrophysics | Harvard \& Smithsonian (CfA), the Chilean Participation Group, the French Participation Group, Instituto de Astrofísica de Canarias, The Johns Hopkins University, Kavli Institute for the Physics and Mathematics of the Universe (IPMU) / University of Tokyo, the Korean Participation Group, Lawrence Berkeley National Laboratory, Leibniz Institut f{\"u}r Astrophysik Potsdam (AIP), Max-Planck-Institut f{\"u}r Astronomie (MPIA Heidelberg), Max-Planck-Institut f{\"u}r Astrophysik (MPA Garching), Max-Planck-Institut f{\"u}r Extraterrestrische Physik (MPE), National Astronomical Observatories of China, New Mexico State University, New York University, University of Notre Dame, Observat{\'o}rio Nacional / MCTI, The Ohio State University, Pennsylvania State University, Shanghai Astronomical Observatory, United Kingdom Participation Group, Universidad Nacional Aut{\'o}noma de M{\'e}xico, University of Arizona, University of Colorado Boulder, University of Oxford, University of Portsmouth, University of Utah, University of Virginia, University of Washington, University of Wisconsin, Vanderbilt University, and Yale University.

\section*{Data Availability}

The data underlying this article will be shared on reasonable request to the corresponding author.

\bibliography{Papers}

\appendix

\section{Redshift evolution} \label{sec:z_evolution}

One of the fundamental assumptions that is made when performing \lya{} reconstructions of QSOs is that there is no significant redshift evolution of the spectral properties of the QSOs. In general, our goal is to predict the \lya{} flux at $z\gtrsim6$ to be able to explore the attenuation of the intrinsic QSO flux by an increasingly neutral IGM. Unfortunately we cannot train our reconstruction pipelines on training samples of similar objects, as they would be equally attenuated, thus we must resort to using QSOs at notably lower redshifts, where we have many more objects. However, in doing so we may bias our intrinsic profile reconstructions if there is significant redshift evolution. For example, we typically construct training sets of QSOs from $2.0 \lesssim z \lesssim 2.5$ and assume these will be representative of the properties of QSOs at $z\gtrsim6$. However, from a large sample of composite spectra spanning $0<z<5$, \citet{Dietrich:2002} observed the line equivalent width of \lya{} to increase with redshift. Within this appendix, we explore this assumption of redshift evolution by breaking down a large sample of QSOs into narrow bins in redshift and reconstructing our blind QSOs using the QSO training set from within each redshift bin. 

Specifically, we use the QSO training sample of Greig et al. (in prep), which selects 30,166 QSOs from the BOSS 16th data release QSO catalogue \citep{Lyke:2020} with a median S/N $>6.5$ spanning $2.0 \lesssim z \lesssim 3.5$. Firstly, to separate out the potential impact of QSO luminosity on the training set (see Appendix~\ref{sec:L_evolution}) we restrict this sample of QSOs to those with a measured luminosity between $45.4 < {\rm log}_{10}(L) < 45.6$ (determined at 1280\AA), which reduces our sample to 9,942 QSOs. Next, we split this sample at fixed luminosity into five separate redshift bins spanning: (i) $2 < z < 2.25$, (ii) $2.25 < z < 2.5$, (iii) $2.5 < z < 2.75$, (iv) $2.75 < z < 2.95$ and (v) $2.95 < z < 3.5$ corresponding to 2384, 3580, 2264, 914 and 783 objects, respectively.

Due to the number of blind QSOs (60) and the different redshift training sets (5), we perform our \lya{} profile predictions based on a nearest neighbour search. For each individual blind QSO, we search for the nearest 50 QSOs in each training set (separated by redshift) that have the closest euclidean distance, $d_{j}$ determined over $1260 < \lambda < 2050$:
\begin{eqnarray} \label{eq:NN}
d_j = \sqrt{\sum_{i}^{\lambda=[1260,2050]} \left[ F_{j}(\lambda_{i}) - F_{\rm blind}(\lambda_{i}) \right]^{2}},
\end{eqnarray}
where $F_{j}$ is the QSO flux for the $j$-th object in the training set and $F_{\rm blind}$ is the target QSO to be reconstructed. The predicted \lya{} flux is then determined from the mean and 68th percentile uncertainty from the 50 nearest neighbours. Note, this nearest neighbour prediction is less accurate than that of the reconstruction pipelines analysed within this work. However, for this study, the actual performance is less relevant, what we actually care about is whether there is a visible trend in the reconstruction profiles for the same object as we train on different QSO samples from different redshifts. 

In order to isolate the impact of redshift evolution on the \lya{} predictions, for each individual blind QSO we take the ratio of the predicted flux in each redshift bin relative to the predicted flux in the lowest redshift bin ($2 < z < 2.25$). If redshift evolution is present, we should observe an increase or decrease in the amplitude of this ratio as we increase the redshift separation relative to the lowest redshift bin. To reduce the amount of information we then take the average of these ratios over the entire sample of 30 blind QSOs for each sample (X-Shooter and SDSS). Since we are measuring the ratio of QSO predictions for each individual object, the spectral properties are removed (i.e. emission lines) and instead when we average we are isolating the smooth differential properties primarily due to the differences in redshifts of the samples.

\begin{figure*} 
	\begin{center}
	  \includegraphics[trim = 0.5cm 0.5cm 0cm 0.5cm, scale = 0.47]{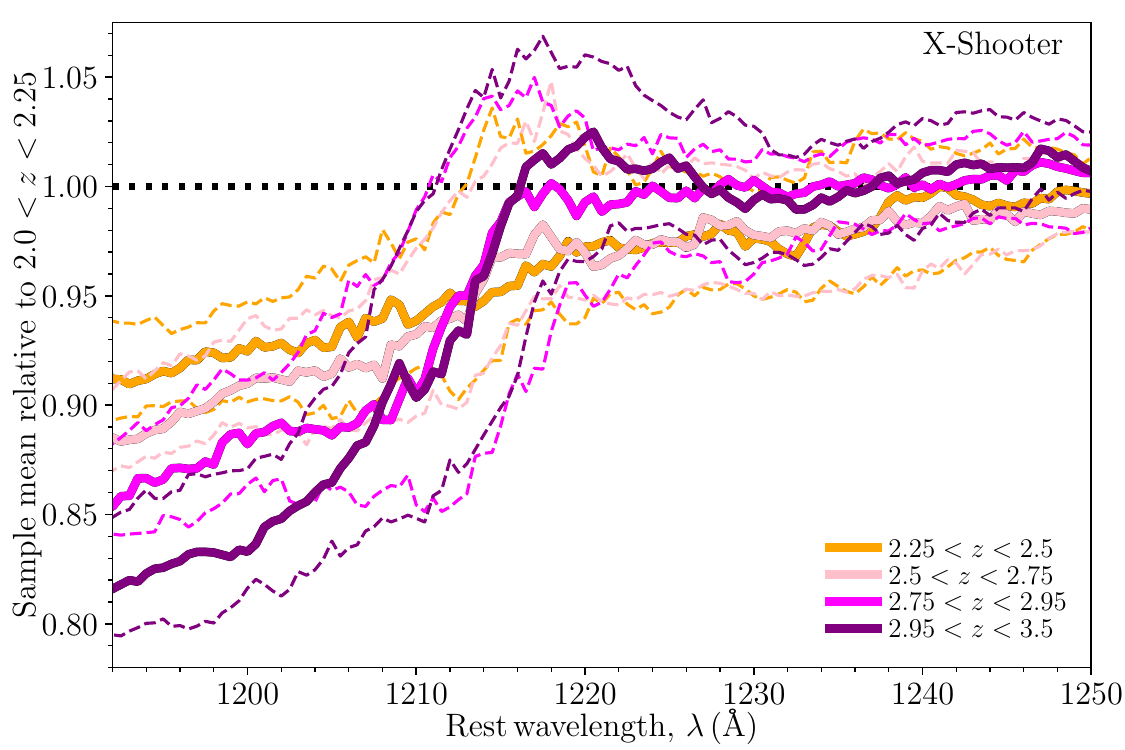}
	  \includegraphics[trim = 0.25cm 0.5cm 0cm 0.5cm, scale = 0.47]{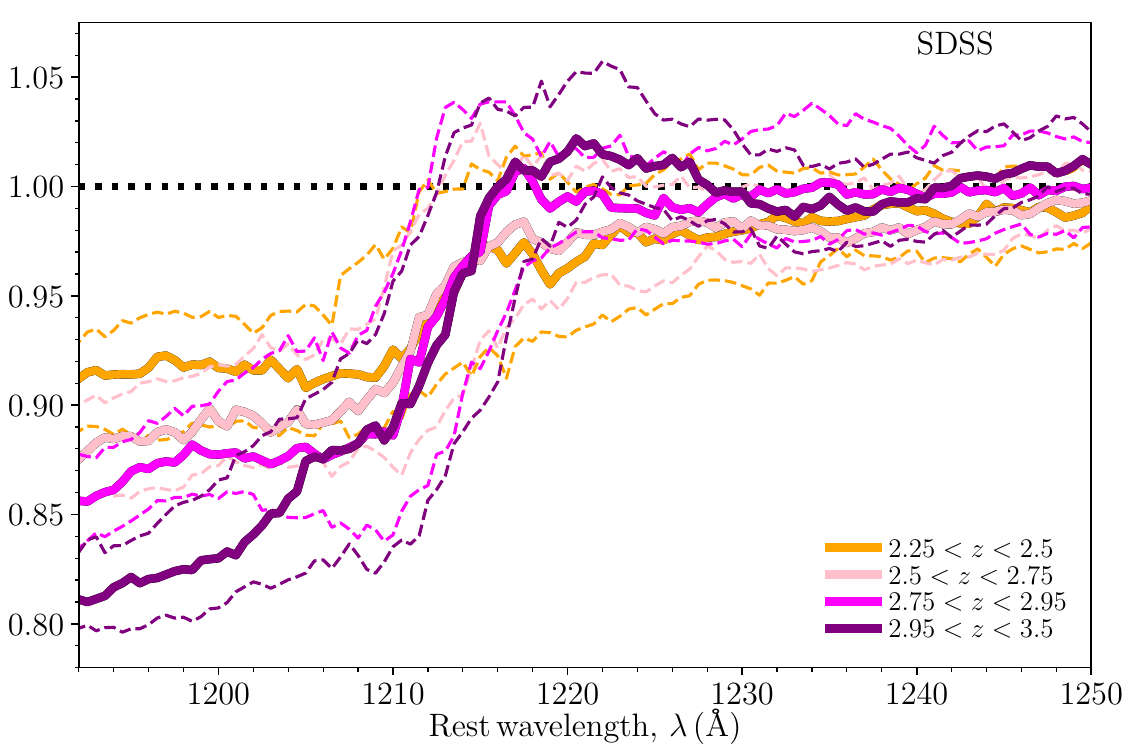}
	\end{center}
\caption[]{The redshift evolution in the mean ratio of predicted QSO flux for fixed QSO luminosity. The mean ratio is determined by taking the mean over the full sample of 30 QSOs in each blind sample of the individual predicted QSO flux ratios obtained from nearest neighbour reconstruction using two different training sets that are separated in redshift (see text for further details). The left panel corresponds to the X-Shooter sample and the right panel corresponds to the SDSS sample. Solid curves correspond to the sample mean while the dashed curves correspond to the 68th percentiles. In all, we compare the flux ratio for increasing redshift bins relative to the lowest redshift bin at $2 < z < 2.25$.}
\label{fig:fixedL}
\end{figure*}

In Figure~\ref{fig:fixedL} we summarise these results, showcasing the sample averaged mean of the predicted QSO flux relative to the first (lowest) redshift bin ($2.0 < z < 2.25$) for the X-Shooter (left panel) and SDSS sample (right panel). The solid curves are the mean and the dashed curves correspond to the 68th percentiles determined over the 30 QSOs within the sample.

Across the two samples the results are essentially identical. In terms of the impact of redshift evolution on the reconstruction profiles we can only use the information redward of \lya{} (i.e. $\lambda > 1215.67$\AA) as blueward of \lya{} is impacted by the increasing attenuation of the \lya{} forest by the IGM with increasing redshift. Figure~\ref{fig:fixedL} clearly demonstrates this, with the sample mean decreasing for increasing redshift of the training set.

Redward of \lya{}, we recover very little evidence for a trend in the mean predicted flux with redshift. Each curve, corresponding to increased redshift bins, are all within $\lesssim5$ per cent of the mean predicted flux relative to the lowest redshift bin. Nevertheless, there is a very weak trend of an increasing QSO flux with increasing redshift (from orange to purple). Note, the amplitude of the mean in the lowest redshift bin is higher than most other bins (ratio is below unity) however the difference is smaller that the corresponding uncertainties thus likely purely random. Although, this is consistent across both samples, but the relative size is extremely small. Ultimately, within the amplitude of the 68th percentile uncertainties drawn across the sample, the results are perfectly consistent with no redshift evolution. Nevertheless, these results are broadly consistent with \citet{Dietrich:2002}. Our redshift range is much narrower than that of \citet{Dietrich:2002}, thus for an equivalent range in their data a weak trend is present, however, within their provided uncertainties it would be equivalent to no evolution. Therefore, to truly determine the validity of assuming no redshift evolution this study needs to be revisted for a larger sample of objects with increased redshift range.

\section{Luminosity evolution} \label{sec:L_evolution}

\citet{Dietrich:2002} also observed that the \lya{} line equivalent width decreased with increasing redshift, with the trend being much stronger than that with respect to redshift evolution. Therefore, we repeat our previous analysis instead fixing our QSO sample by redshift and binning by luminosity. Using the same base QSO sample as previously, we first restrict it to include only QSOs within $2.25 < z < 2.5$, corresponding to 10,551 QSOs. We then split this sample into five luminosity bins spanning $44.5 < {\rm log}_{10}(L) < 45.3$, $45.3 < {\rm log}_{10}(L) < 45.5$, $45.5 < {\rm log}_{10}(L) < 45.7$, $45.7 < {\rm log}_{10}(L) < 45.85$ and $45.85 < {\rm log}_{10}(L) < 46.5$, which corresponds to 1572, 3915, 2800, 1144 and 1114 QSOs, respectively.

\begin{figure*} 
	\begin{center}
	  \includegraphics[trim = 0.5cm 0cm 0cm 0.5cm, scale = 0.47]{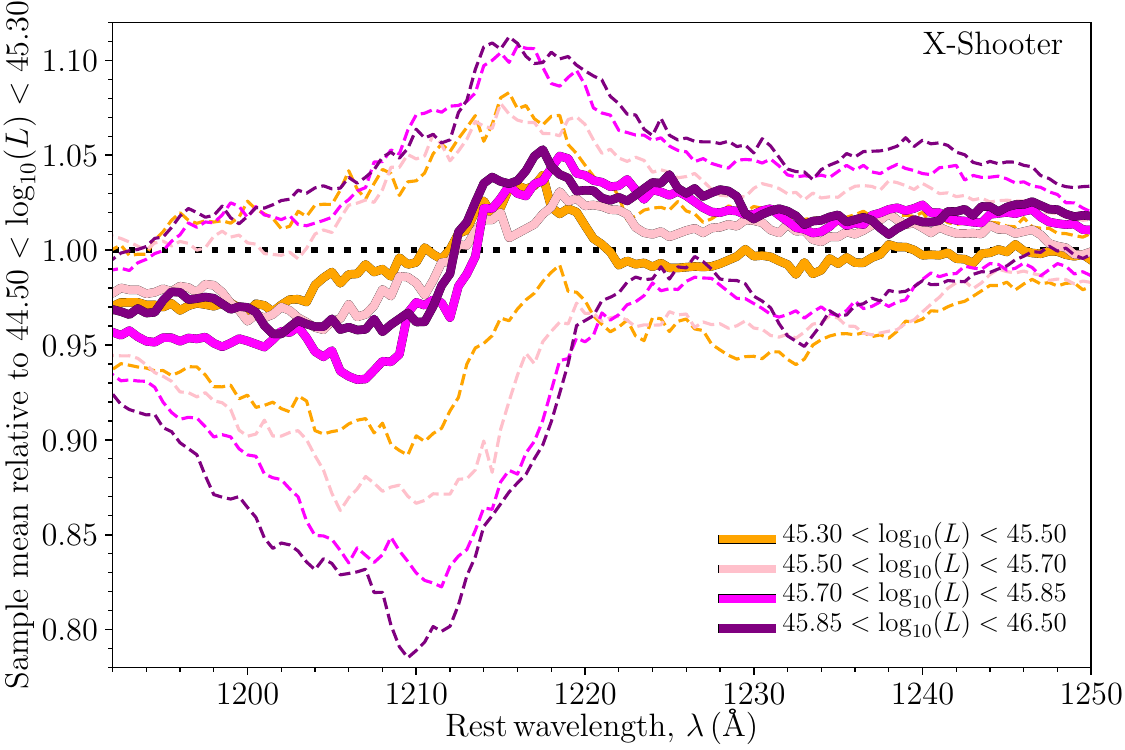}
	  \includegraphics[trim = 0.25cm 0cm 0cm 0.5cm, scale = 0.47]{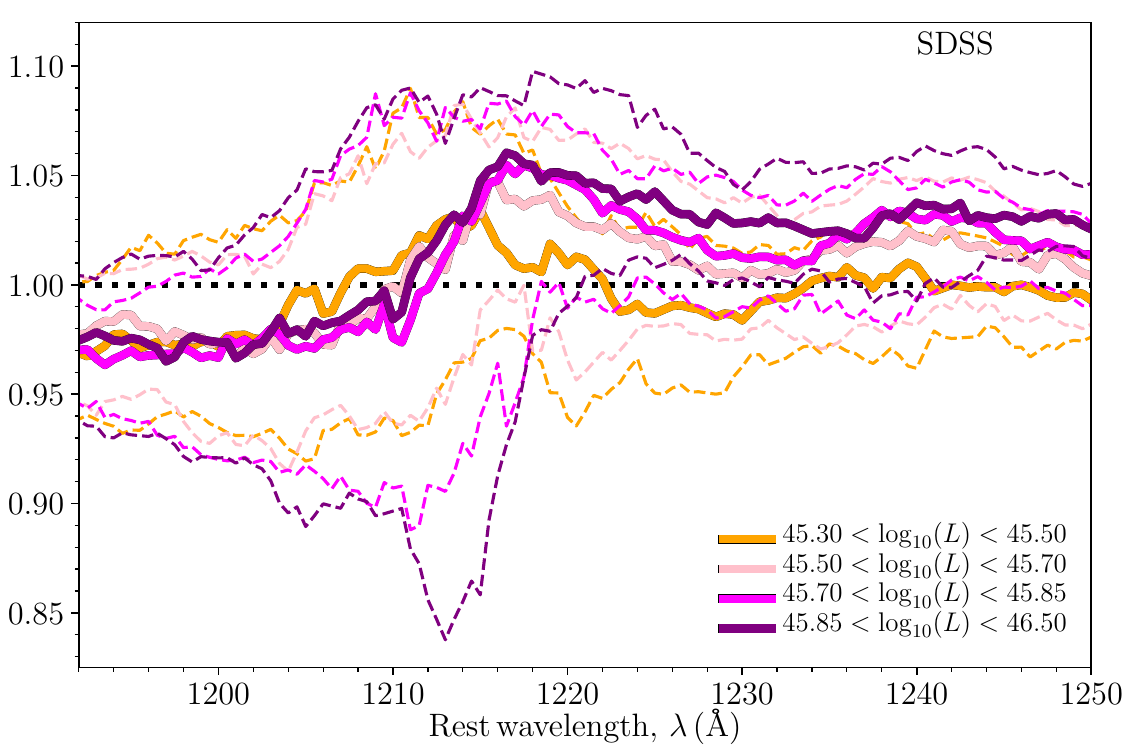}
	\end{center}
\caption[]{The same as Figure~\ref{fig:fixedL} except now we consider the impact of luminosity evolution for fixed redshift. Here, we present the evolution relative to the lowest luminosity bin, measured at 1280\AA, of $44.5 < {\rm log}_{10}(L) < 45.3$.}
\label{fig:fixedz}
\end{figure*}

In Figure~\ref{fig:fixedz} we summarise the results of the sample mean determined from the individual ratios of predicted \lya{} flux relative to the predicted \lya{} flux from the lowest luminosity bin ($44.5 < {\rm log}_{10}(L) < 45.3$).   Again, the left panel corresponds to the X-Shooter sample of blind QSOs and the right panel corresponds to the SDSS sample. As we are considering a fixed redshift range, we are unaffected by increasing IGM attenuation in the \lya{} forest and thus can consider the full spectral coverage red and blueward of \lya{}.

Once again the results are consistent across both samples and we appear to recover a weak trend of an increasing predicted \lya{} flux for increasing luminosity. However, again the relative amplitude of this increase, over $\sim1.5$ dex in luminosity, is $\sim5$ per cent, well within the much broader statistical uncertainties. Therefore, once again we conclude that there is no significant evolution of the \lya{} line properties with intrinsic QSO luminosity. 

This appears inconsistent with the strong trend of decreasing equivalent width with luminosity as observed by \citet{Dietrich:2002}. However, we note that: firstly, the trend observed by these authors is with respect to equivalent width and not flux as a function of wavelength as within our study. Secondly, the properties of the blind QSOs within each sample vary quite broadly (i.e. containing both weak and strong \lya{} line profiles) which correspond to vastly different equivalent widths. The composite samples constructed by \citet{Dietrich:2002} are preferentially selected by their strong \lya{} properties (i.e. large equivalent widths). Therefore in constructing the mean relation over our entire blind sample, we are in effect averaging out the effect with equivalent width. However, if we instead focus on the 68th percentiles from our sample means, in particular the lower 16th percentiles (i.e. lower dotted curves), the relative amplitude increasingly deviates from unity for an increasing luminosity. This implies an increasingly larger reduction in the QSO flux for some QSOs with increasing luminosity, consistent with a decreasing equivalent width for increasing luminosity. Therefore, if we were to select those QSOs from our blind sample as representatives of the QSOs that entered into the \citet{Dietrich:2002} composites, we would more clearly observe this trend. This implies that care may need to be taken to construct a training set consistent with the luminosity of the target QSO to be reconstructed. However, the relative amplitude of the effect is $\lesssim10-15$ per cent and further the line strengths of the other redward emission lines tends to reduce to a similar extent with increasing luminosity as \lya{}. Therefore, it is unlikely that evolution with QSO luminosity plays a significant role.

However, once again the number of QSOs within this study is relatively small along with the width of explored QSO luminosities. Therefore, one should consider repeating this analysis with a much larger sample of objects.

% Don't change these lines
\bsp	% typesetting comment
\label{lastpage}
\end{document}